\documentclass[sigconf]{acmart}
\AtBeginDocument{%
  }

\setcopyright{acmlicensed}
\copyrightyear{2026}
\acmYear{2026}
\acmDOI{XXXXXXX.XXXXXXX}
\acmConference[MAB'26]{Make sure to enter the correct
  conference title from your rights confirmation email}{September 21--22,
  2026}{Bangkok, TH}
\acmISBN{978-1-4503-XXXX-X/2026/09}

\usepackage{tabularx}
\usepackage{booktabs}
\usepackage{array}
\usepackage{graphicx}
\usepackage{pifont}
\usepackage{supertabular}
\usepackage{float}
\begin{document}

\title{The Evolution and ‘Chineseness’ of Urban Screens in China: From Propaganda Posters to Smart City Interfaces}

\author{Wuwu Ran}
\authornote{First and corresponding author.}
\email{wuwu.ran@york.ac.uk}
\orcid{0009-0001-0238-5149}
\affiliation{%
    \institution{University of York}
    \city{York}
    \country{United Kingdom}
}

\renewcommand{\shortauthors}{Wuwu Ran}

\begin{abstract}
Urban screens have become a global feature of the media city, yet their developmental paths are not universal. This paper argues that the evolution of urban screens in China constitutes a distinct paradigm of media urbanism, termed Chineseness, understood here not as cultural essence but as a relational, historically situated configuration of three linked elements: deep-rooted lineages of public communication; state-steered technological leapfrogging driven by industrial policy and mega-events; and a regulatory paradox that enables spectacular scale while tightening operational control. Drawing on a five-stage genealogy (1990–present), a chronology of representative screens, and five paradigmatic cases, the study shows how this configuration produces temporal–spatial compression, the coexistence of multiple technological generations, and privileges collective witnessing over device-centric interactivity. The paper contributes a documented genealogy of Chinese urban screens, develops temporal–spatial compression and collective witnessing as analytical tools for non-Western media ecologies, and positions Chineseness as a globally relevant alternative to market-led models of media architecture.

\end{abstract}

\ccsdesc[500]{Human-centered computing~Interaction design}
\ccsdesc[500]{Applied computing~Media arts}
\ccsdesc[500]{Applied computing~Architecture (buildings)}
\ccsdesc[300]{Social and professional topics~Governmental regulations}

\keywords{Urban Screens, Media Architecture, \textit{Chineseness}, Urban Screen Clusters, Augmented Public Space}

\received{13 June 2025}
\received[revised]{13 September 2026}
\received[accepted]{21 September 2026}

\maketitle

\section{Introduction}
Urban screens have become a defining feature of the global media city, but their paths of development are not universal. China offers a distinctive, accelerated trajectory. Rapid, state-led urbanisation and targeted industrial policy, exemplified by the `Hundreds of Cities, Thousands of Screens' UHD initiative\footnote{A national digital display initiative launched as a five-year plan (2019--2023), reissued in 2021 following pandemic disruptions, to promote ultra-high-definition (UHD) infrastructure and smart-city screen networks across China. Referred to hereafter as the \textit{UHD initiative}.}, have supported a nationwide network of landmark displays and a domestic LED sector with a significant global share\cite{wang2011role,hu2015influence,sanderson2014light,ye2018evolution,coema2001,Xing2023}. Despite this scale, academic discourse remains skewed toward Western cases and single-site studies in China, leaving the broader history and governance logics undertheorised \cite{mcquire2006politics,colangelo2015expanded,berry2013shanghai,schielke2021media}. This paper addresses that gap by proposing \textit{Chineseness} as a situated configuration that shapes the evolution of urban screens. This study argues that the evolution of urban screens in China forms a distinct paradigm of media urbanism, \textit{Chineseness}, defined by three linked elements: lineages of public communication (from dazibao and open-air cinema to architectural illumination)\cite{huhtamo2009messages,turnock2015plastic,berry2013shanghai}; state-driven technological leapfrogging driven by industrial policy and mega-events\cite{butollo2018great,lauermann2019urban}; and a durable enable-and-constrain regime that scales spectacle while tightening day-to-day control. Using a five-stage genealogy (1990–present) and paradigmatic cases, this paper shows how this configuration yields temporal–spatial compression and privileges collective witnessing over individualized interactivity. This research contributes a compact genealogy, two operational concepts, and a specification of \textit{Chineseness} as a globally relevant alternative to market-led models.

\section{Methods}
\subsection{Research Design and Questions}
This study employs an explanatory, qualitative historical, and comparative design that links macro documentary analysis with interviews and site-based inquiry. Three questions guide the analysis. RQ1 (Genealogy): How have Chinese urban screens evolved since the 1990s, and which recurrent drivers mark inflexion points? RQ2 (Governance/\textit{Chineseness}): How do industrial policy, municipal regulation, and operator practice co-produce a regime that scales spectacle while constraining content? RQ3 (Publicness/Curation): How do flagship screens coordinate collective witnessing and operator-as-curator practices, and how do these differ from Western paradigms?

\subsection{Historical and Case Comparison}
The historical review examines urban screen development in China from 1990 to 2024, drawing on national and municipal policies and standards, industry documents, planning and architectural materials, archival sources, and historical accounts. The factor map aligns macro-environmental shifts, regulation, mega-events, and screen-technology development along a shared timeline to trace concurrent changes and temporal lags. The resulting five stages constitute an interpretive periodisation of these intersecting developments. They describe shifts in emphasis rather than a uniform sequence followed by every city or the complete replacement of one technology by another.

The study compares People's Square, Aurora Plaza, GreenPix, One Shenzhen Bay, and 3788 Light of Asia. These five cases were purposively selected to capture variation across historical stages, urban settings, screen types, and operating arrangements. \textit{Light of Asia} receives more detailed ethnographic analysis, while the remaining four cases are examined as stage exemplars within the comparative framework presented in Table~1.

\subsection{Sources and Empirical Scope}
The research triangulated four corpora: policy and industry documents (directives, standards, nightscape guidelines, and LED/UHD initiatives); media-archaeological materials (planning files, architectural documents, patents, photographs, videos, and project sheets); site-based visual ethnography involving repeated day- and night-time observations of spatial integration, crowd behaviour, and programming rhythms; and platform traces and paratexts such as operator calendars, festival and PSA schedules, and widely circulated clips of 3D sets (Appendices A.1, A.4, A.5).  The interview corpus contains ten participants interviewed in Mandarin during 2022–2023, with excerpts translated into English. Participants include architects, planning professionals and officials, a digital media strategist, a manufacturer representative, and visitors with experience of the relevant urban settings. Appendix A.2 provides participant profiles and selected excerpts.

\section{A Five-Stage Genealogy (1990–Present) and Cases}

\begin{figure*}[t]
    \centering
    \includegraphics[width=\textwidth]{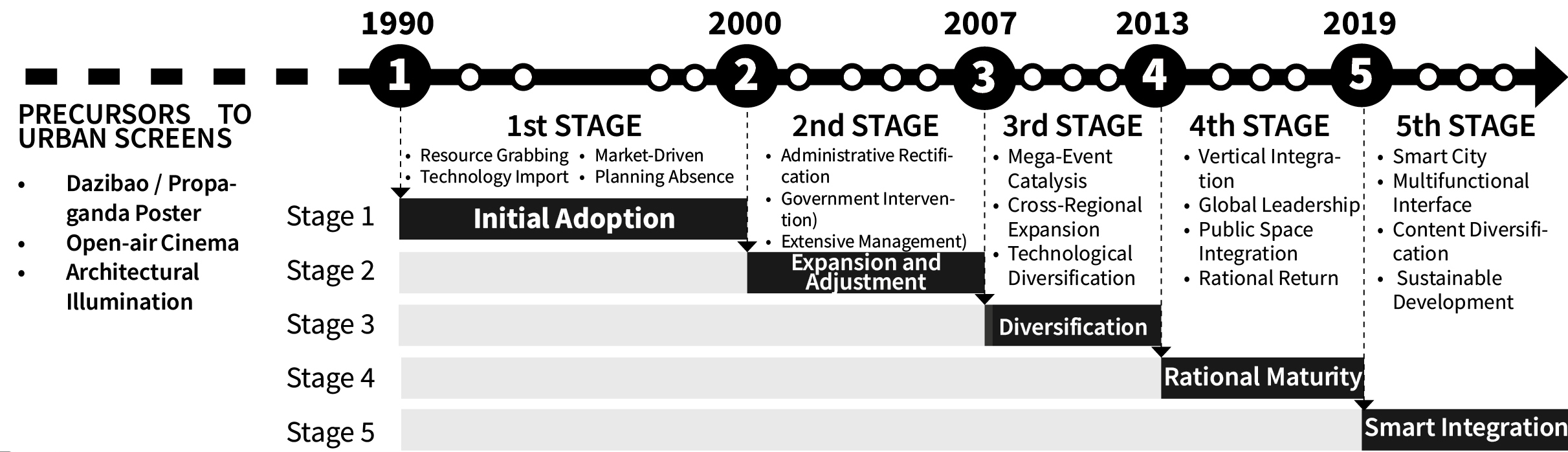}
    \caption{Five-stage evolution of Chinese urban screens
    (1990--present). Diagram by the author.}
    \label{stage}
\end{figure*}

\subsection{A Short History of Urban Screens Evolution in China (1990–Present)}
Drawing on the chronology in Appendix A.1 and supporting corpora, the factor map in Appendix A.3 historicises the transformation of media urbanism in China through four co-evolving forces: macro-environmental shifts, regulatory reforms, mega-event hosting, and screen-technology development. Urbanisation and GDP growth aligned with industrial policy initiatives such as the National Semiconductor Lighting Project (2003) and the 13th/14th Five-Year Plans. Olympic and Expo cycles legitimised experimental façades, while municipal codes on luminance, curfews, and siting stabilised their operation. These interactions delineate five evolutionary stages (1.0–5.0) in Chinese urban screen development from 1990 to 2024 (Fig. 1). Their boundaries identify shifts in emphasis, while earlier display forms and operating practices may persist. The cases that follow examine how these broad developments take different forms in particular projects.

Stage 1.0 (1990s): Initial Adoption. Large LED displays emerged amid early marketisation and rapid urban growth. Post-reform policies spurred real-estate development creating new façades and plazas. Breakthroughs in full-colour blue LED technology and the rapid growth of domestic securities markets from the early 1990s, which spurred demand for large display screens in trading halls \cite{coema2001}, catalysed a shift from static lightboxes and CRTs to large LED billboards that redefined urban visibility.

Stage 2.0 (2001-2006): Expansion and Adjustment. China's real-estate boom drove brand-oriented urbanism where illuminated skylines projected cosmopolitan modernity \cite{gu2014large}. Local governments competed for investment and visibility within a centralised framework \cite{xu2011chinas}, yet rapid proliferation prompted municipal 'clean-up' campaigns removing thousands of non-compliant screens \cite{wang2011role}. This set in place the enable-and-constrain logic: curbing excess while endorsing flagship façades. The National Semiconductor Lighting Project (2003) laid industrial groundwork \cite{ye2018evolution,butollo2018great} as cities experimented with siting regulations.

Stage 3.0 (2007-2012): Diversification (Mega-Event-Driven). Mega-events, exemplified by the Beijing Olympics (2008), accelerated urban transformation and recast cityscapes as national showcases \cite{shin2012unequal,wu2016reproducing}. Since 2001, over sixty international events advanced an upgrade-by-event strategy \cite{ma2024domestic}, legitimising experimental media architectures. Urban screens became nation-branding instruments employing façades, interactive prototypes, and art commissions to project modernity and sustainability to global audiences \cite{deng2013conceptualizing,kobierecki2021sports}.

Stage 4.0 (2013-2018): Rational Maturity. Urban screens matured into planned public-space infrastructure. Small-pitch LED advances enabled high-resolution façades and district-scale control, shifting from singular spectacle to ambient, data-responsive choreography. Citywide shows (starting with Nanchang's 2013 light show) turned skylines into 'massive media' \cite{10.1145/2946803.2946812}: continuous urban canvases for state-orchestrated spectacle \cite{schielke2021media}. Governance shifted to enforceable standards on luminance, content, and maintenance, aligning with industry vertical integration and 'green' policies \cite{zhang2023fuzzy,chen2022critical}.

Stage 5.0 (2019-Present): Smart Integration. Top-down UHD initiative  and viral spectacle pursuits integrate 8K resolution, cloud scheduling, sensor networks, and 3D anamorphic screens creating depth illusions \cite{wang2024paintings}. By 2021, nearly 100 projects signalled a shift from billboard logic to immersive, shareable displays (2022 China LED Industry White Paper). Isomorphic escalation (companies competing for larger, higher-density screens) drives operators to curate original content. Screens now function as curated cultural platforms and smart-city interfaces, blending advertising with culturally resonant IP, seasonal rituals, and AR-based participation.

\subsection{Case Studies of Five Stages}
\subsubsection{Stage 1.0 — People’s Square Screen (Shanghai, 1994–2002)}
\subparagraph{\textit{The People's Square Screen} shows how programming and location gave early display technology a lasting social role. Inspired by a temporary screen installed in the square for the 1990 Beijing Asian Games, it linked mega-event experimentation to permanent urban infrastructure (Appendix~A.1, Figure~\ref{guanyinqiaoA9}). Broadcasts of news, music, civic announcements, advertising, and live sports supported collective viewing~\cite{mcquire2017geomedia}, echoing earlier traditions of communal viewing and social gathering around open-air cinema in China~\cite{Li2022}. Its visibility made `Meet under the Large Screen' a familiar meeting instruction before widespread mobile communication (Appendix~A.2). Although redevelopment removed the screen in 2002, memories persisted: a brand's 2018 campaign at Raffles City appropriated this meeting ritual to promote a pop-up store. The case shows how repeated social use can give screens cultural significance that outlasts their physical presence and becomes a resource for commercial promotion (Fig.~2).}

\begin{figure}[htbp]
    \centering
    \includegraphics[width=\linewidth]{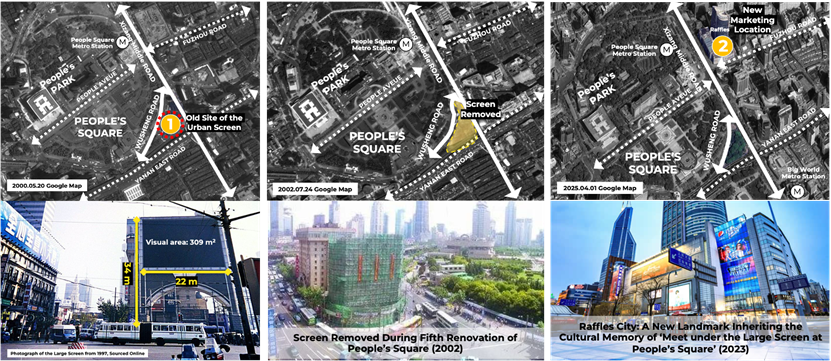}
    \caption{The People's Square urban screen
    (Shanghai, 1994--2002).
    Composite figure and analytical annotations by the author.
    Base map: Google Maps.
    Photographs sourced from
    \href{https://sghexport.shobserver.com/html/toutiao/2020/10/26/285402.html}
    {Shangguan News}
    and
    \href{https://www.redhotmedia.com.cn/outdoor_info_67530.html}
    {redhotmedia.com.cn}.}
    \label{fig:peoples-square}
    \Description{Composite figure combining a Google Maps
    base map, the author's analytical annotations, and
    photographs of the People's Square urban screen in Shanghai.}
\end{figure}

\subsubsection{Stage 2.0 — Aurora Plaza (Shanghai, 2003)}

\subparagraph{\textit{Aurora Plaza} established a prominent advertising surface within the Lujiazui--Bund nightscape~\cite{gu2014large,lin2014shanghai}. Across the Huangpu River, its low-resolution LED matrix forms coherent imagery, making viewing distance central to its visual effect. Commercial advertising shares the screen with official city promotions and holiday broadcasts (Appendix~A.2). The borrowed `I~$\heartsuit$~SH' slogan packages civic attachment for global consumption, raising questions about authenticity, intended audiences, and the treatment of Shanghai's visual heritage~\cite{pawlik2020illuminating,lin2014shanghai}. Interviewees link its advertising premium to landmark status and describe planning restrictions as protecting the nightscape while preserving scarcity (Appendix~A.2). Its reliance on advertising and familiar branding formulas also exposes tensions between promoting a globally recognisable city image and representing the complexity of local culture. The case illustrates the selective enabling and regulation of urban screens: city promotion reinforces symbolic value, while planning controls limit further expansion. It exemplifies a model of media facades in China, combining programmable imagery, skyline visibility, commercial value, and municipal approval (Fig.~3).}

\begin{figure}[htbp]
    \centering
    \includegraphics[width=\linewidth]{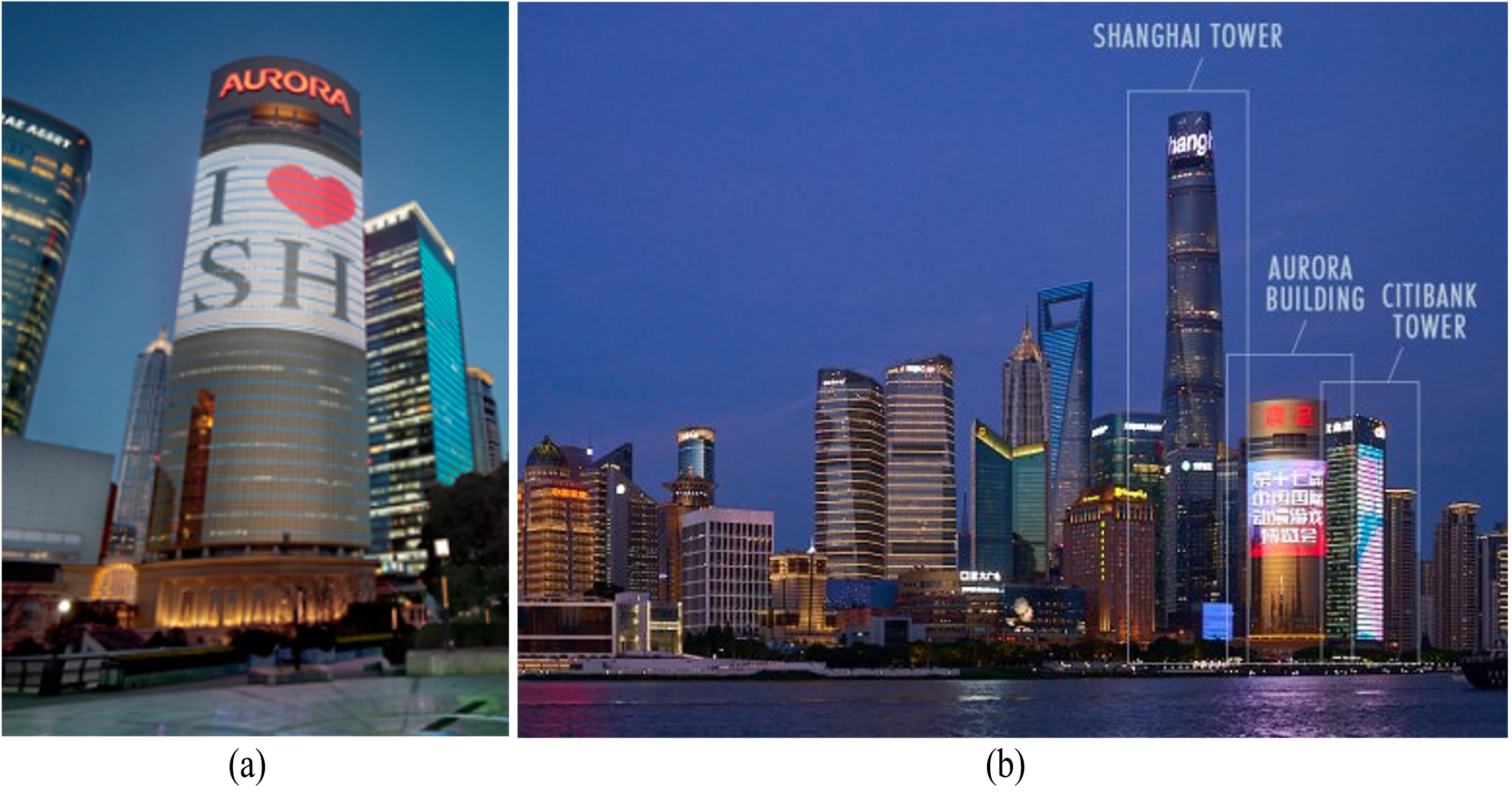}
    \caption{Aurora Plaza and the Lujiazui nightscape.
    (a) The media facade on Aurora Plaza displaying
    'I\ding{170} SH';
    (b) the wider Lujiazui skyline.
    Photographs sourced from
    \href{https://www.smartshanghai.com/articles/shanghai-life/the-explainer-this-is-how-you-buy-advertising-on-the-pudong-skyline}
    {SmartShanghai}
    and
    \href{https://m.jiemian.com/article/686217.html}
    {Jiemian News}.}
    \label{fig:aurora}
    \Description{Two photographs showing Aurora Plaza's
    illuminated media facade displaying an I love Shanghai
    message and a wider view of the Lujiazui skyline at night.}
\end{figure}

\subsubsection{Stage 3.0 — GreenPix Zero-Energy Media Wall (Beijing, 2008)}

\subparagraph{The \textit{GreenPix} Zero-Energy Media Wall integrated photovoltaic cells and RGB LEDs into a curtain wall designed for daytime energy harvesting and low-resolution nocturnal imagery. Olympic visibility legitimised experimentation, while art and promotional broadcasts expressed green, high-tech, and people-centred ambitions~\cite{giostra2011fachada,xu2011perspective}.  Its relevance to the comparison lies in how environmental discourse and media art accompanied a moment of heightened international visibility. Costs and complexity raised replication concerns; limited interactivity left it primarily a broadcast surface. Its zero-energy claim invites scrutiny of operational and lifecycle performance through critiques of performative greenness~\cite{stevanovic2010use, perini2014sustainable}. Nevertheless, GreenPix became canonical by showing how environmental discourse and media art could legitimise civic screens beyond advertising, linking ecological rhetoric with national representation (Fig.~4).}

\begin{figure}[htbp]
    \centering
    \includegraphics[width=\linewidth]{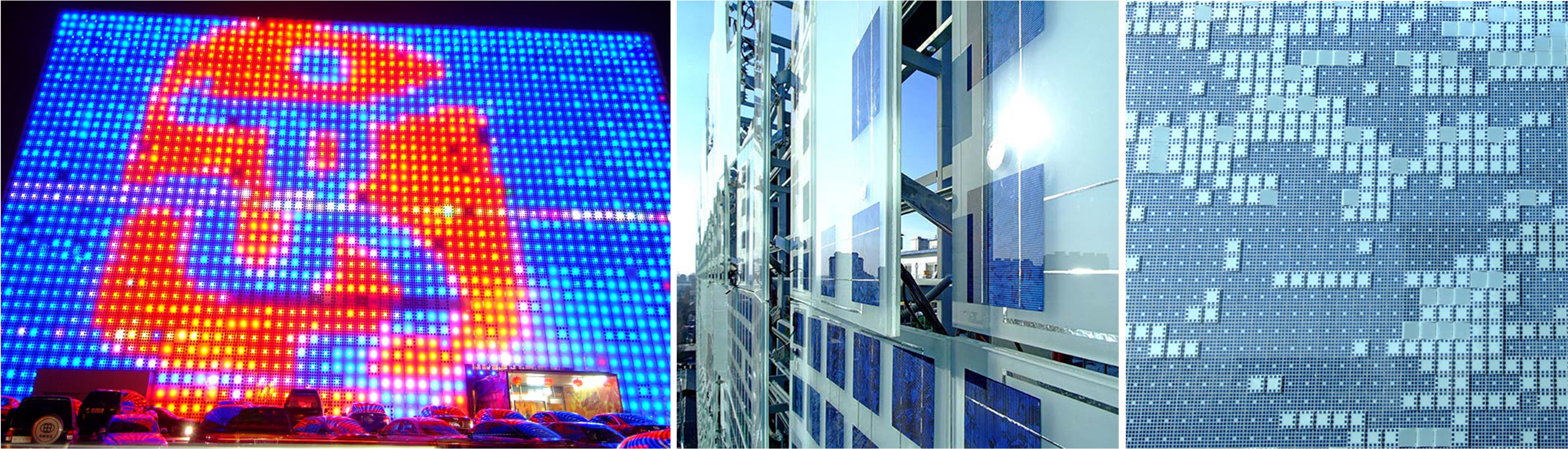}
    \caption{GreenPix Zero Energy Media Wall at the
    Xicui Entertainment Complex, Beijing.
    Image source:
    \href{https://www.archdaily.com/zh/624598/greenpix-ling-hao-neng-mei-ti-qiang-e-fay-jones}{ArchDaily}.}
    \label{fig:greenpix}
    \Description{The GreenPix media wall at the
    Xicui Entertainment Complex in Beijing.}
\end{figure}

\subsubsection{Stage 4.0 — One Shenzhen Bay (2018)}
\subparagraph {\textit{One Shenzhen Bay} 's tower media facade marks a shift towards rational maturity, from singular icons to data-responsive infrastructure. It recasts the facade as a calibrated environmental interface rather than a continuously illuminated sign. Carefully positioned LED lines minimise glare and maintain distant legibility without disturbing occupants or dominating the skyline~\cite{licht2021}. On-site environmental data animate the imagery, creating a poetic dialogue between screen and ecosystem. Shenzhen's regulatory innovation supports this approach. A proposed walking programme would convert residents' steps into microdonations and integrate anonymised mobility data into the facade's narrative~\cite{zheng2021}; its implementation is not documented in the case materials. The project exemplifies the Stage~4.0 configuration: high-resolution, sensor-responsive, and policy-compliant media architecture that prioritises ecological responsiveness and operational stewardship over sheer brightness or continuous advertising (Fig.~5).}
\begin{figure}[htbp]
    \centering
    \includegraphics[width=\linewidth]{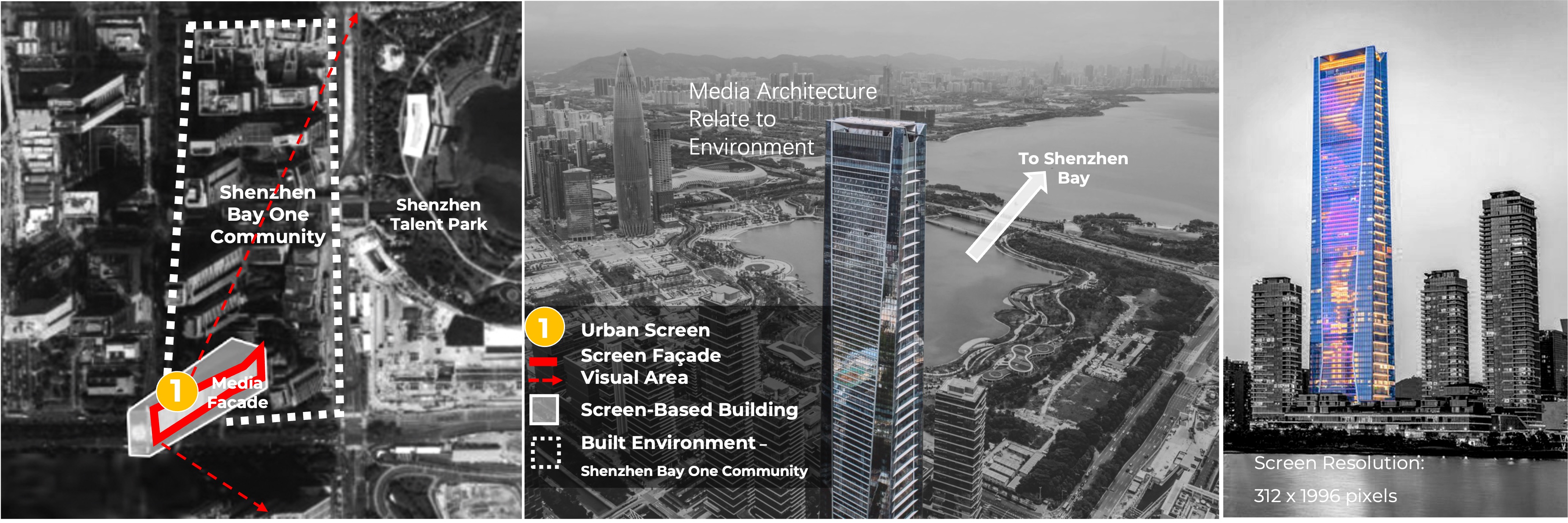}
    \caption{One Shenzhen Bay and its media fa\c{c}ade. Composite figure and analytical annotations by the
author. Base map: Google Maps.
    Photo credit: Shenzhen Parkland Real Estate Group
    Co., Ltd. Source:
    \href{https://www.lichtkunstlicht.com/projects/one-shenzhen-bay/}{Licht Kunst Licht}.}
    \label{fig:one-shenzhen-bay}
    \Description{One Shenzhen Bay in Shenzhen, showing
    the illuminated media fa\c{c}ade.}
\end{figure}

\subsubsection{Stage 5.0 — Light of Asia (2019)}

\subparagraph{\textit{Light of Asia} combines a 3,788~m$^2$ LED facade with 8K anamorphic imagery and an operator-as-curator model. Following the pandemic-related advertising downturn, DIGMEGA expanded its role from selling screen space to producing cultural programmes. Works such as \textit{Track Across Building}, \textit{Sweater for the Screen}, and \textit{Super Moon + AR Constellations} connect local imagery, seasonal rhythms, and mobile participation (Appendix~A.5). Spaceflight broadcasts organise collective attention around national achievements, extending the logic of ceremonial media events into urban space~\cite{dayan1992media}. Advertising remains predominant, but artistic commissions and event programming broaden the screen's public role. Viewers' photography and online sharing carry these encounters beyond the square, giving local imagery and media events national and global reach~\cite{dayan1992media,michael2002public}. Through these practices, the screen shifts from background ambience to a platform for what McQuire terms \textit{collective witnessing}~\cite{mcquire2017geomedia}, transforming the square into an augmented public space. The case exemplifies Stage~5.0 through the combination of immersive imagery, cultural curation, and networked circulation. It positions the screen as both medium and curator: a programmable urban interface through which operators orchestrate civic emotion, policy messaging, and viral spectacle within a governance-first smart-city framework, reinforcing local identity and urban visibility (Fig.~6).}
\begin{figure}[htbp]
    \centering
    \includegraphics[width=\linewidth]{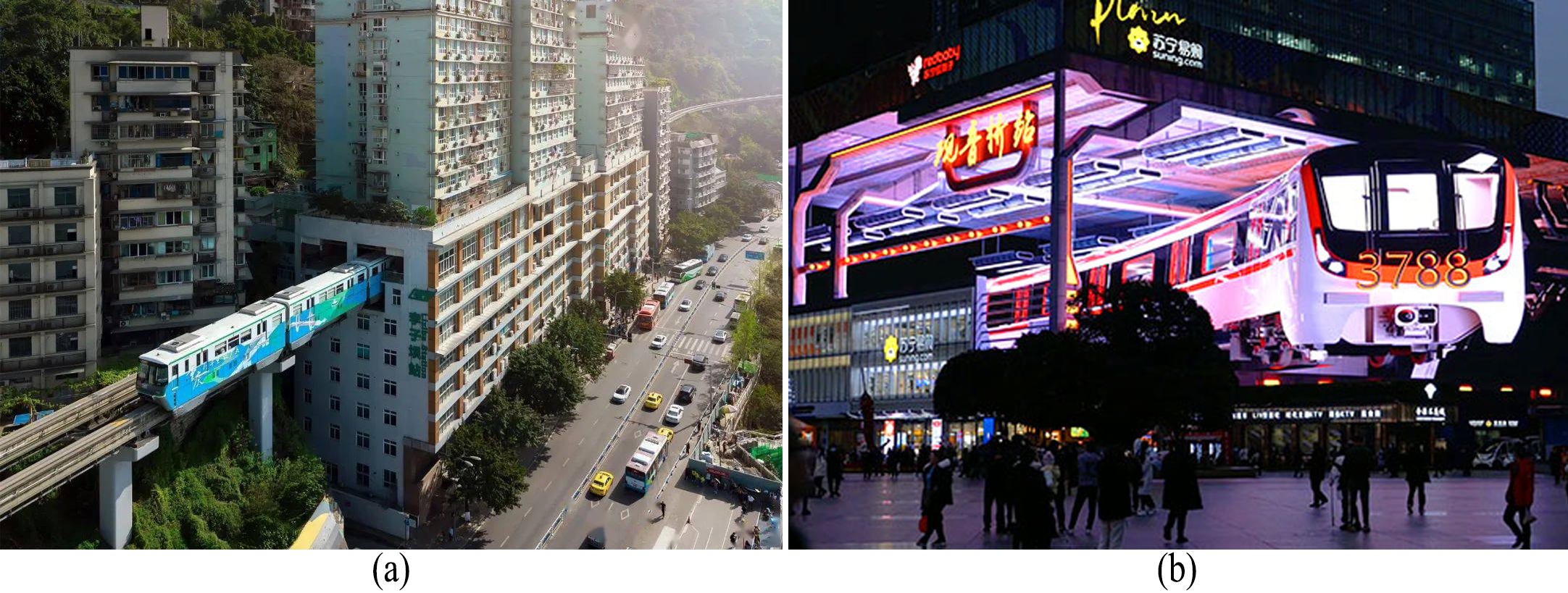}
    \caption{(a) Liziba, Chongqing. Photo: Zhang Kunkun,
    via \href{https://www.sohu.com/a/494911032_121106884}{Sohu}.
    (b) \textit{Track Across Building (HELLO, CHONGQING!)} by VISFIELD.
    Image: \href{https://museaward.com/winner-info.php?id=223236}
    {MUSE Creative Awards}.}
    \label{fig:liziba-hello-chongqing}
    \Description{(a) A train passing through a building
    at Liziba. (b) A digital train appearing to emerge
    from a building-mounted 3D screen.}
\end{figure}
\subsection{Comparative Summary}
\begin{table*}[t]
    \caption{Comparative overview of five cases aligned
    to the five-stage genealogy}
    \label{tab:freq}
    \centering
    \small

    \setlength{\tabcolsep}{4pt}
    \renewcommand{\arraystretch}{1.15}

    \newcommand{\MABcaseimage}[1]{%
        \raisebox{-\height}{%
            \includegraphics[
                width=\linewidth,
                height=1.8cm,
                keepaspectratio
            ]{#1}%
        }%
    }

    \begin{tabular}{
        @{}
        >{\raggedright\arraybackslash}p{2.1cm}
        *{5}{
            >{\raggedright\arraybackslash}
            p{\dimexpr
                \textwidth/5-2.1cm/5-2\tabcolsep
            \relax}
        }
        @{}
    }
        \toprule

        \textbf{Project}
        & \textbf{People's Square Screen}
        & \textbf{AURORA Plaza}
        & \textbf{GreenPix}
        & \textbf{One Shenzhen Bay}
        & \textbf{3788 Light of Asia} \\

        \midrule

        Stage
        & Stage 1.0
        & Stage 2.0
        & Stage 3.0
        & Stage 4.0
        & Stage 5.0 \\

        \addlinespace[3pt]

        Image
        & \MABcaseimage{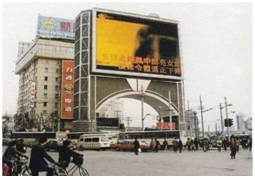}
        & \MABcaseimage{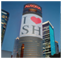}
        & \MABcaseimage{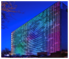}
        & \MABcaseimage{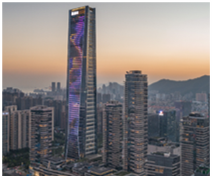}
        & \MABcaseimage{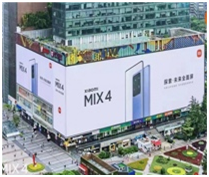} \\

        \addlinespace[4pt]

        Time
        & 1994--2002
        & 2003
        & 2008
        & 2018
        & 2019 \\

        Location
        & Shanghai
        & Shanghai
        & Beijing
        & Shenzhen
        & Chongqing \\

        Type (MAB)
        & (3)
        & (1)
        & (2)
        & (4)
        & (5) \\

        \midrule

        Design / operation
        & Sino-Japanese collaboration
        & Nikken Sekkei; Taiwan Glass Ind. Corp.
        & Simone Giostra \& Partners and ARUP
        & Kohn Pedersen Fox/LKL
        & Operator: DIGMEGA \\

        \addlinespace[4pt]

        Purpose
        & First large public LED.
        & Skyline branding.
        & Olympic-era innovation.
        & Environmental choreography.
        & City branding and tourism. \\

        \addlinespace[4pt]

        Screen attributes
        & Central square; 309~m$^2$; ``outdoor TV''.
        & High-rise media fa\c{c}ade; ``I \ding{170} SH''.
        & Zero-energy PV-LED hybrid wall; 2,200~m$^2$.
        & Sensor-responsive lines; glare control.
        & 3D anamorphic; AR; 3,788~m$^2$. \\

        \addlinespace[4pt]

        Primary beneficiaries
        & Public (meeting point/information).
        & City image; tourists; advertisers.
        & Public; cultural sector; government.
        & Residents; planners; policy makers.
        & Public; businesses; tourists. \\

        \addlinespace[4pt]

        Relation to public space
        & Central crossroads (People's Square).
        & Iconic skyline position (Bund--Lujiazui).
        & Street-facing fa\c{c}ade near Olympic venues.
        & Integrated into the new Bay Area development.
        & L-shaped wrap on a pedestrian-level building. \\

        \bottomrule
    \end{tabular}

    \par\smallskip

    \parbox{\textwidth}{%
        \footnotesize
        \raggedright
        \textit{Note.}
        Type categories follow the five-fold classification
        of media architecture defined by the Media Architecture
        Biennale (MAB)~\cite{mab_categories}:
        (1) ``Money'' and entertainment architecture;
        (2) Animated architecture;
        (3) Infrastructure and participatory architecture;
        (4) Spatial media art;
        (5) Future trends and prototypes.
    }
\end{table*}
Read across Table~1, the five cases position \textit{Chineseness} as a historically situated paradigm of media urbanism: a configuration in which national development priorities, municipal discretion, commercial interests, and cultural practices jointly shape urban display. The comparison examines three interlocking dimensions: public-communication lineages, state-steered technological leapfrogging, and the enable-and-constrain regulatory paradox. The cases provide complementary evidence of these relationships, whose relative importance changes across the five stages.

First, public-communication lineages persist through changes in the organisation of attention. Comparing People's Square with Light of Asia shifts the focus from display capabilities to how public encounters are assembled and circulated. Both connect everyday urban experience with civic and national narratives, but the relationship between gathering, participation, and commercial mediation changes. Collective witnessing thus remains a recurrent practice while its organisation and spatial reach evolve~\cite{mcquire2017geomedia}. This comparison places cultural continuity in the repeated use and reinterpretation of screens, while tracing a shift towards the commercial curation of encounters across physical and digital space.

Second, technological leapfrogging describes selective acceleration within an uneven developmental landscape. The factor map (Appendix~A.3) connects opportunities for experimentation to industrial policy, mega-event visibility, and expanding technological capacity. Research on China's LED industry situates this capacity within interactions among local state support, domestic demand, and global production~\cite{butollo2018great}. Across GreenPix and the later cases, however, technical sophistication does not entail equivalent advances in participation or environmental performance. Leapfrogging is therefore uneven: advanced hardware can coexist with established communicative practices and experimental operating models. GreenPix's international collaboration further shows how nationally situated ambitions can be pursued through transnational expertise.

Third, the enable-and-constrain paradox identifies governance as a process that produces value as well as setting limits. Aurora and One Shenzhen Bay reveal different priorities in Shanghai and Shenzhen: protecting landmark visibility and commercial scarcity in one setting, and moderating environmental effects and integrating display with building use in another. These differences are consistent with China's combination of national political direction and substantial local initiative~\cite{xu2011chinas}. Municipal discretion helps determine which displays gain prominence and what conditions sustain their operation. Light of Asia adds the operator's role in translating civic and national narratives into programming that also attracts audiences and supports commercial activity. Governance and markets consequently interact through decisions about location, design, and content, shaping whose interests and experiences receive visibility. Publicness is thus organised selectively: access to shared imagery can expand without equivalent opportunities to influence what is shown.

Together, the cases locate Chinese specificity in the historical articulation of these relationships. Neither advanced hardware nor shared viewing alone establishes \textit{Chineseness}; the explanatory contribution lies in connecting screen practices to China's institutional arrangements and cultural histories. The comparison shows how a nationally directed but locally differentiated development context shapes the purposes, beneficiaries, and public roles of urban screens. It also qualifies the stage model: technical acceleration can coexist with persistent communicative practices, and new capabilities need not displace earlier forms. Section~4 examines how this uneven development materialises spatially within urban screen clusters.

\section{URBAN SCREEN CLUSTER – SPATIAL COMPRESSION AND AUGMENTED PUBLIC SPACE}

Urban screen clusters represent a spatial manifestation of China's media urbanism. As Melzer notes, screens concentrate in ``highly orchestrated and strategically selected spaces''~\cite{melzer2019screen}, generating cumulative effects. In China, policy coordination amplifies this tendency, shifting the analytical focus from individual facades to the screen cluster as a meso-scale unit where governance, technology, and collective spectatorship converge. This perspective makes it possible to examine relationships among installations, pedestrian settings, and programming.

Guanyinqiao exemplifies this transformation. Selected from a database of over 100 clusters~\cite{ran2023machine}, the Guanyinqiao cluster comprises seven buildings with screens integrated into a network of pedestrian plazas and walkways (Appendix~A.4, Fig.~A.4). Anchored by Light of Asia's 3,788~m$^2$ corner screen (77.1), which features 3D anamorphic content, the cluster includes shopping mall facades (77.2--77.6) and high-rise office screens (77.7), with display technologies ranging from conventional LED displays to programmable curtain walls. The inventory documents technological and spatial heterogeneity: 8K displays operate metres from decade-old billboards within a coordinated visual field. Temporal--spatial compression offers an interpretation of this juxtaposition within the wider historical account.

\begin{figure}[htbp]
    \centering
    \includegraphics[width=\linewidth]{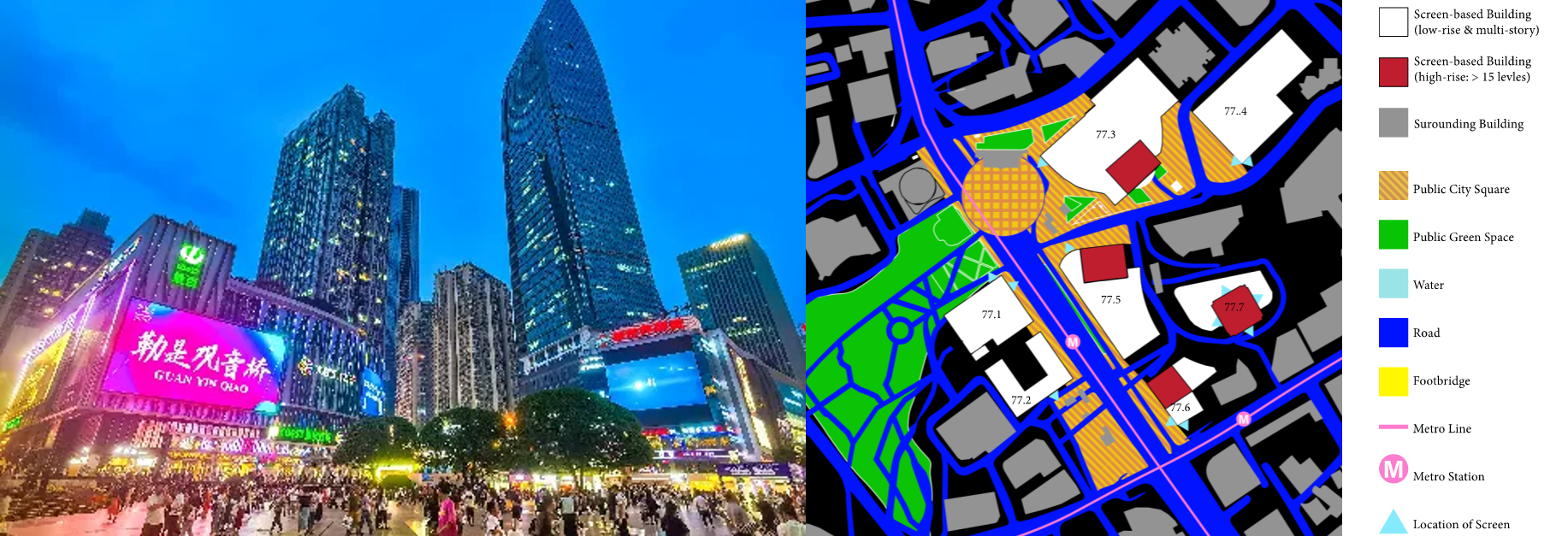}
    \caption{The Guanyinqiao screen cluster.
    Left: photograph provided by the Publicity Department
    of the CPC Jiangbei District Committee, via
    \href{https://www.sohu.com/a/850212888_121106884}{Sohu}.
    Right: morphological mapping by the author.}
    \label{fig:guanyinqiao-morphology}
    \Description{Left: illuminated screens and crowds
    in Guanyinqiao. Right: a map showing screen-based
    buildings, public spaces, roads, and transport facilities.}
\end{figure}

The programming inventory and P1's account of viewing and sharing show how content can extend the significance of an installation beyond its physical surface (Appendices~A.2 and~A.5). AR and social circulation provide additional contexts in which screen content is encountered. \textit{Augmented public space} here describes this relationship between physical settings and mediated encounters. Its quality is assessed not only through fixed architectural attributes but also through the capacity to coordinate attention and sustain a collective audience over time through dynamic layers of information.

\section{DISCUSSION}

This study demonstrates how China's urban screens constitute a distinct paradigm through three enduring patterns: lineages of public communication that enable collective witnessing; state-steered technological leapfrogging that produces temporal–spatial compression; and an enable-and-constrain regime that expands spectacular infrastructures while tightening operational control. These patterns extend scholarship on media cities \cite{mediacity}, civic screens \cite{struppek2005}, programmable façades \cite{haeusler2009media}, massive media \cite{colangelo2019building,10.1145/2946803.2946812}, situational publics \cite{krajina2014urbanising}, and curated platforms \cite{verhoeff2012mobile,melzer2019screen,reinhard2016making}.

China's pathway diverges from Western narratives in three respects. First, governance-first development. Where Western models often frame screens through market pluralism or civic contestation, China's trajectory prioritises state coordination. People's Square (Stage 1.0) programmed national sporting events for community-building; Light of Asia (Stage 5.0), while commercially operated, coordinates major screens with municipal festivals and broadcasts authority-approved content. Industrial policy enables existence; local regulations calibrate operation. The same infrastructures that extend perception also coordinate schedules, manage public mood, and deliver civic messages.

Second, temporal–spatial compression through punctuated upgrades. State-steered leapfrogging creates landscapes where technological generations coexist. In Guanyinqiao, the hyper-advanced 3D anamorphic Light of Asia operates metres from decade-old LED billboards: a heterogeneity produced by event-driven upgrades rather than linear replacement. This compressed environment privileges collective witnessing: People's Square succeeded by gathering crowds for shared, synchronised experiences; Light of Asia perfects this for the 21st century with spectacles like the 3D monorail designed for simultaneous plaza viewing and online sharing. This choreographed participation, rooted in China's open-air cinema as ideological outreach and now elaborated as urban media events \cite{mcquire2017geomedia}, resembles but is not identical to Benjamin's notion of ‘collective reception’ \cite{benjamin2008workofart}. Its didactic lineage contrasts with Western models that privilege personalised, device-centric interactivity.

Third, curated platform logic. Operational models evolved from public utility to hybrid platforms. People's Square operated as a public broadcaster; Light of Asia functions as a broadcast-platform hybrid, strategically producing viral cultural IP that builds brand value and increases commercial advertising rates. This curatorial model reflects state-capitalist urbanism: industrial policy builds hardware, municipal rules manage externalities, and operators translate both into viable, audience-focused cultural products.

These differences can be explained. Event cycles and industrial policy (semiconductor lighting programmes, the Olympics and Expo) produce leaps that outpace bottom-up diffusion models. An enable-and-constrain design (subsidies and UHD initiatives paired with siting limits, content vetting, and brightness standards) organises investment while preserving message discipline. Finally, a distinct operator ecology, often vertically integrated with manufacturers, assumes curatorial responsibility for cadence, seasonal theming, and cross-media IP, a role less typical in market-led environments.

Several findings were unexpected. First, textual address persisted (a semiotic lineage from poster cultures, seen in tickers and slogans) despite richer visual affordances. Second, operators showed curatorial resistance within constraint, using local lore, seasonal rituals, and playful illusions to reframe official or commercial messages without breaching rules, which points to a degree of micro-autonomy under tight oversight. Third, sustainability remained ambivalent: Stage 4's data-responsive façades suggest environmental responsiveness, yet Stage 5's UHD scaling raises unresolved questions about energy use and light trespass. Future work should measure operational energy and luminance over seasons and separate those findings from lifecycle impacts rather than assume 'smart = green' \cite{zhang2023fuzzy,chen2022critical}.

Limitations: Cases skew toward first-tier cities; methods are qualitative, lacking longitudinal energy, luminance, and attention traces. Future work should extend multi-sited ethnography across city tiers, pair programming logs with sensor-based measurement, deploy computational content analysis, and pursue comparisons with Seoul, Tokyo, and Dubai to test the portability of these analytical tools.

\section{CONCLUSION}

This paper traced Chinese urban screens from their 1990s emergence to today's smart spectacles, revealing how five stages culminate in a distinct paradigm: \textit{Chineseness}. This configuration integrates deep-rooted lineages of public communication, state-steered technological leapfrogging, and enable-and-constrain governance into a coherent historical logic. Chinese urban screens emerge as historically situated infrastructures of collective attention where innovation and regulation evolve in tandem.

Theoretically, this evidence-based genealogy extends media-architecture scholarship by foregrounding temporal–spatial compression and collective witnessing as analytical tools for non-Western media ecologies, offering a critical counterpoint to interaction-centric models that dominate scholarship. Practically, it offers guidance for integrating digital layers (screens, programmable façades) into architectural planning from the outset rather than as afterthoughts; for regulators, fostering micro-autonomy in curatorial practice while preserving night-sky quality and content oversight; for operators, investing in curatorial capacity rather than pixel arms races.

As cities worldwide embed digital media into the urban fabric, the Chinese case is instructive, though its lessons are mixed. It shows how compelling, lasting urban spectacle emerges from the interplay of state ambition, regulatory precision, and curatorial authorship. The future of augmented public space lies precisely at this intersection, where orchestrated collectivity meets orchestrated control.

\begin{acks}
This article draws on several chapters of the author's
doctoral thesis. The author thanks their doctoral
supervisors, Professor Richard Koeck and
Dr Francesca Piazzoni, for their guidance, support,
and valuable feedback throughout the doctoral research.
\end{acks}

\bibliographystyle{ACM-Reference-Format}
\bibliography{reference}

\appendix
\section{APPENDIX}
\subsection{The Chronology of Urban Screens}

\par
\begingroup
\small
\setlength{\tabcolsep}{3pt}
\renewcommand{\arraystretch}{1.12}
\newcommand{\ChronThumb}[1]{%
    \raisebox{\dimexpr\ht\strutbox-\height\relax}{%
        \includegraphics[width=1.10cm,height=1.10cm,keepaspectratio]{#1}%
    }%
}
\topcaption{Chronology of representative urban screens (1988-2024).\label{table1}}
\tablefirsthead{%
    \toprule
    \textbf{Year} & \textbf{Location and description} & \textbf{Image} \\
    \midrule
}
\tablehead{%
    \multicolumn{3}{@{}l@{}}{\small\itshape Table~\ref{table1} (continued)} \\[3pt]
    \toprule
    \textbf{Year} & \textbf{Location and description} & \textbf{Image} \\
    \midrule
}
\tabletail{\bottomrule}
\tablelasttail{\bottomrule}
\begin{supertabular}{@{}
    >{\raggedright\arraybackslash}p{0.72cm}
    >{\raggedright\arraybackslash}p{\dimexpr\columnwidth-1.90cm-4\tabcolsep\relax}
    >{\centering\arraybackslash}p{1.18cm}
    @{}}
        1988 & \textbf{Beijing}\newline
            Beijing Train Station square. Outdoor LED screen at the railway station in the city of Beijing (visual area: $12\,\mathrm{m} \times 10\,\mathrm{m}$).
        & \ChronThumb{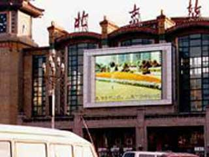} \\
        \addlinespace[4pt]
        \shrinkheight{4pt}
        1990 & \textbf{Shanghai}\newline
            The Foundation for First Urban Screen (People’s Square) in Shanghai Installed a $7.2 \text{m} \times 10.8\text{m}$ colour television display at People's Square to broadcast the Asian Games.
        & {} \\
        \addlinespace[4pt]
        \shrinkheight{4pt}
        1993 & \textbf{Beijing}\newline
            First Colour LED Screen in Beijing. The inaugural outdoor colour advertising screen in Beijing, boasting a visual area of $25\text{m}^2$, was officially inaugurated.
        & {} \\
        \addlinespace[4pt]
        \shrinkheight{4pt}
        1994 & \textbf{Shanghai}\newline
            First Urban Screen (People’s Square) in Shanghai. Shanghai's First Screen at People's Square, composed of over 5.39 million LEDs, could display text, animations, and broadcast TV programs.
        & \ChronThumb{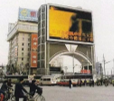} \\
        \addlinespace[4pt]
        \shrinkheight{4pt}
        1998 & \textbf{Zhengzhou}\newline
            Outdoor Led Screen at Zhengzhou’s Railway Station (visual area: $16 \text{m} \times 13.5\text{m}$)
        & \ChronThumb{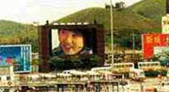} \\
        \addlinespace[4pt]
        \shrinkheight{4pt}
        1998 & \textbf{Shenzhen}\newline
            Outdoor Led Screen at Shenzhen’s Railway Station (visual area: $12 \text{m} \times 9\text{m}$)
        & \ChronThumb{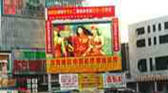} \\
        \addlinespace[4pt]
        \shrinkheight{4pt}
        1999 & \textbf{Chongqing}\newline
            Asian largest Screen. One of the World’s Largest Outdoor Led Screen in Chongqing (visual area: $39 \text{m} \times 22\text{m}$).
        & \ChronThumb{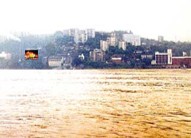} \\
        \addlinespace[4pt]
        \shrinkheight{4pt}

        2003 & \textbf{Shanghai}\newline
            AURORA Plaza LED Façades. The AURORA International Building LED façades was built and opened in 2003 and was awarded the Guinness World Record for the largest screen in October 2004. Visual area: $57 \text{m} \times 63\text{m}$.
        & \ChronThumb{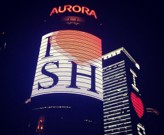} \\
        \addlinespace[4pt]
        \shrinkheight{4pt}
        2005 & \textbf{Guangzhou}\newline
            Guangzhou Tower (Yuexiu).The 'National Cellar(Guojiao) 1573' advertisement on the body of Guangzhou TV Tower (Yuexiu Mountain) was once China's largest outdoor neon advertisement in 2005. This advertisement was removed in October 2009.
        & \ChronThumb{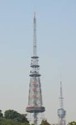} \\
        \addlinespace[4pt]
        \shrinkheight{4pt}
        2005 & \textbf{Shanghai}\newline
            Oriental Mall, Shanghai. The China’s first curved advertising and commercial screen.
        & \ChronThumb{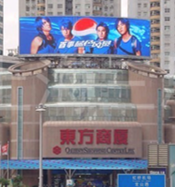} \\
        \addlinespace[4pt]
        \shrinkheight{4pt}
        2006 & \textbf{Shanghai}\newline
            The Pentagon Plaza. The Egg, Bailianyouyicheng Mall and Wanda Plaza, where they are embedded in a huge range of LED screen.
        & \ChronThumb{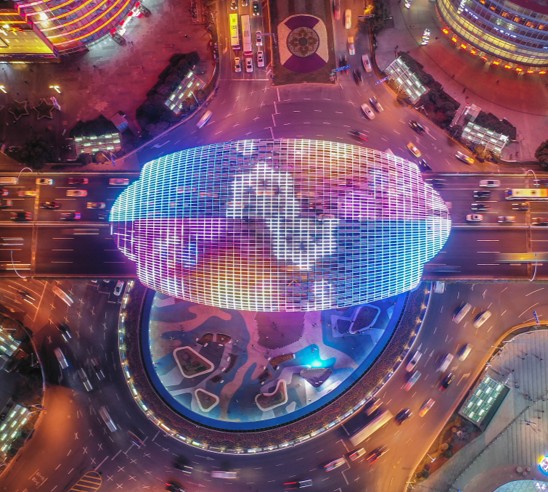} \\
        \addlinespace[4pt]
        \shrinkheight{4pt}
        2007 & \textbf{Beijing}\newline
            The Place, Canopy Screen. A huge LED screen was installed in the main block of Beijing’s Central Business District. visual area: $250 \text{m} \times 30\text{m}$. And 80 feet high.
        & \ChronThumb{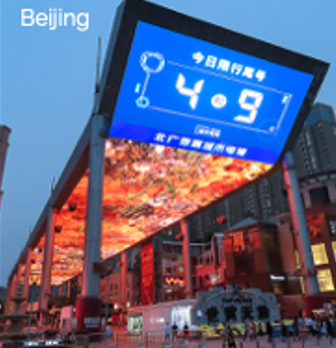} \\
        \addlinespace[4pt]
        \shrinkheight{4pt}

        2008 & \textbf{Beijing}\newline
            GreenPix Zero-Energy Media Wall. China's first photovoltaic media façade, on the curtain wall of the Xicui Entertainment Complex (locally known today as the Moli Hotel), near the 2008 Olympic venues. Designed by Simone Giostra \& Partners with Arup. The display comprises 2,292 RGB LED points over a 2,200 m\textsuperscript{2} (24,000 sq ft) media wall integrated into a building. Solar cells laminated in the glass curtain wall power the screen at night.
        & \ChronThumb{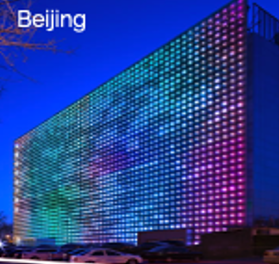} \\
        \addlinespace[4pt]
        \shrinkheight{4pt}
        2008 & \textbf{Beijing}\newline
            Sanlitun Village. The façade of the multifunctional club ‘Orange Hall’ is a $25 \text{m} \times 9\text{m}$. A large screen in a small, enclosed, commercial plaza.
        & \ChronThumb{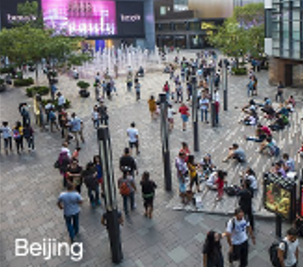} \\
        \addlinespace[4pt]
        \shrinkheight{4pt}
        2009 & \textbf{Beijing}\newline
            Tiananmen Square, Beijing. The Most Expensive Public Service Screens in the world. The two giant LEDs at Tiananmen Square in Bejing. Each with Visual area: $2000\,\mathrm{m}^2$
        & \ChronThumb{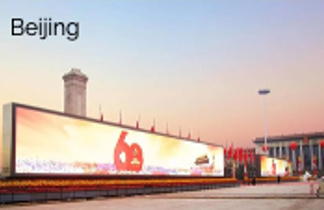} \\
        \addlinespace[4pt]
        \shrinkheight{4pt}
        2009 & \textbf{Suzhou}\newline
            Harmony TimesSquare, Suzhou. The LED canopy is the longest and largest screen area in the world, with a visual area $500 \text{m} \times 32\text{m}$.
        & \ChronThumb{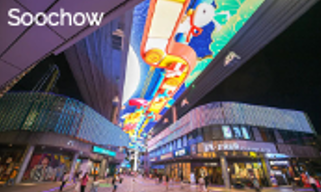} \\
        \addlinespace[4pt]
        \shrinkheight{4pt}
        2009 & \textbf{Guangzhou}\newline
            Canton Tower, Guangzhou. Guangzhou New TV Tower, is the tallest tower in China. The main body of the tower is 454 metres with a total height of 600 meters. Visual area: $133 \text{m} \times 51\text{m}$.
        & \ChronThumb{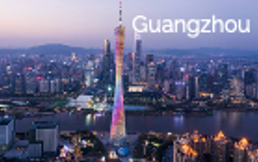} \\
        \addlinespace[4pt]
        \shrinkheight{4pt}

        2010 & \textbf{Shanghai}\newline
            Shanghai Expo Axis. It comprised six valleys, each 50 meters high and 80 meters in diameter, incorporating 200,000 LED modules and over 30,000 glass panels capable of displaying 16 million colour variations.
        & \ChronThumb{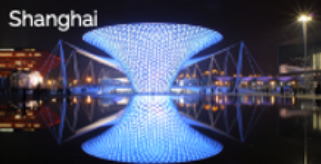} \\
        \addlinespace[4pt]
        \shrinkheight{4pt}
        
2012 & \textbf{Shenzhen}\newline
Wongtee Plaza, Shenzhen. A large outdoor advertising
screen extends along the shopping-centre frontage
and wraps around its corner.
& \ChronThumb{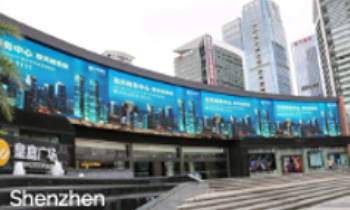} \\
\addlinespace[4pt]
\shrinkheight{4pt}
        2013 & \textbf{Nanchang}\newline
            First Light Show in China, Nanchang. The Nanchang Riverfront Landscape Lighting Renovation Project has featured simultaneous light shows across 293 buildings. Its spectacular illuminated belt extends for 8 kilometers along both sides of the river.
        & \ChronThumb{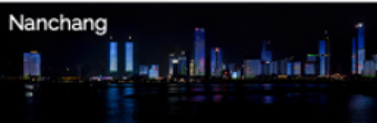} \\
        \addlinespace[4pt]
        \shrinkheight{4pt}
        2013 & \textbf{Wuhan}\newline
            Han Street Wanda Plaza, Wuhan. Wanda Plaza features what was then the world's largest LED control system, comprising 3.18 million LED bulbs. It also incorporates the world's only double-layer media façade.
        & \ChronThumb{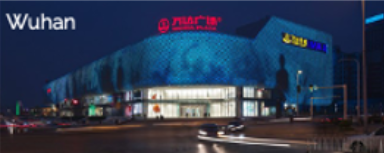} \\
        \addlinespace[4pt]
        \shrinkheight{4pt}
   2014 & \textbf{Shenzhen}\newline
Fortune Building, Shenzhen. A high-definition outdoor
advertising screen extends across three sides of the
building, measuring approximately 180~m in length
with a display area of nearly $800\,\mathrm{m}^2$.
& \ChronThumb{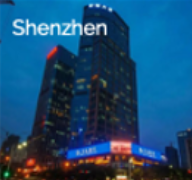} \\
\addlinespace[4pt]
\shrinkheight{4pt}
        2014 & \textbf{Wuhan}\newline
            Han Show Theatre, Wuhan. The theatre‘s design draws inspiration from the traditional Chinese paper lantern, realized on a monumental scale. Its surface is adorned with 18,000 discs and features a programmable LED façade. visual area: $3300\,\mathrm{m}^2$=$30 \text{m} \times 110\text{m}$)
        & \ChronThumb{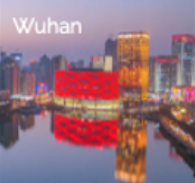} \\
        \addlinespace[4pt]
        \shrinkheight{4pt}

        2015 & \textbf{Chongqing}\newline
            Maoye Tower. Asia's first outdoor LED screen lit up Chongqing Maoye Tower, the whole screen and the building's two walls fusion, an area of 3,363 square metres, a height of more than 100 metres.
        & \ChronThumb{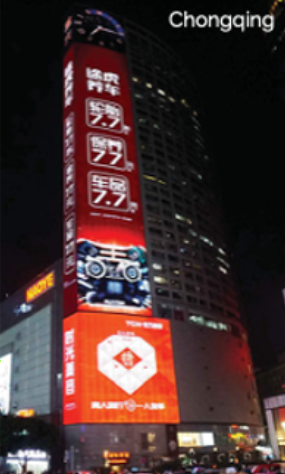} \\
        \addlinespace[4pt]
        \shrinkheight{4pt}
        2016 & \textbf{Hangzhou}\newline
            G20 Summit, Hangzhou. China's technological expertise shines through Qianjiang New City's programmable LED façade project, encompassing 35 buildings with 360,000 LED light points, 5,800 wall washers and floodlights, and 8,200 transparent light guide plates. This pioneering domestic initiative implemented synchronized intelligent lighting control via a 4G wireless private network with real-time reporting and remote debugging capabilities.
        & \ChronThumb{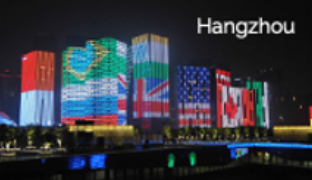} \\
        \addlinespace[4pt]
        \shrinkheight{4pt}
        2017 & \textbf{Shanghai}\newline
            The Aegean Plaza. The Aegean Place is composed of an integrated, programmable LED façade seamlessly incorporated into the building's architecture, complemented by a substantial mounted LED display panel measuring 259.2 square meters. This media architecture designed by a meticulously crafted aluminium panel exterior, enhanced with customised, integrated, and controllable LED light.
        & \ChronThumb{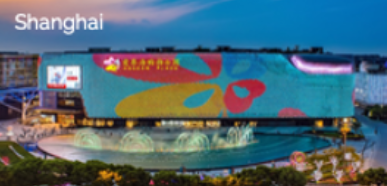} \\
        \addlinespace[4pt]
        \shrinkheight{4pt}
        2018 & \textbf{Shenzhen}\newline
            One Shenzhen Bay, Shenzhen. The building façade project uses real-time environmental data to create media art that connects humans, nature, and architecture, raising environmental awareness.
        & \ChronThumb{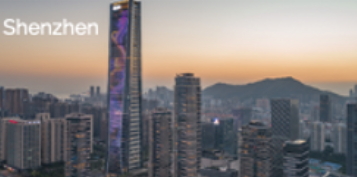} \\
        \addlinespace[4pt]
        \shrinkheight{4pt}
        2018 & \textbf{Chongqing}\newline
            Wings of Chongqing, Jiefangbei. LED screen display on a commercial building in Jiefangbei Pedestrian Street. (visual area: 426 m\textsuperscript{2} = $31.2 \text{m} \times 13.68\text{m}$)
        & \ChronThumb{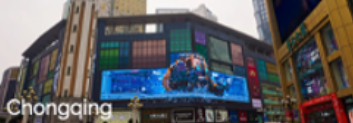} \\
        \addlinespace[4pt]
        \shrinkheight{4pt}

        2019 & \textbf{Zhengzhou}\newline
            Haishane Plaze, Zhengzhou. The large LED canopy and an outdoor grille screen use ETFE for high light transmission and weather resistance, creating an iconic commercial landmark.
        & \ChronThumb{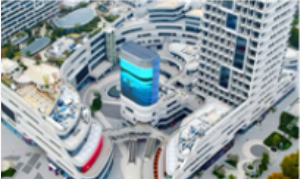} \\
        \addlinespace[4pt]
        \shrinkheight{4pt}
        2019 & \textbf{Chongqing}\newline
            3788 the Light of Asia. The first giant screen in Asia, known as the light of Asia, is Naked-Eye 3D display in Guanyinqiao, Chongqing.
        & \ChronThumb{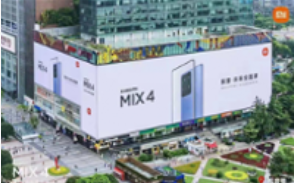} \\
        \addlinespace[4pt]
        \shrinkheight{4pt}
        2020 & \textbf{Shanghai}\newline
            The mirror hall of the Pudong Art Museum consists of two spaces, both 55m long, with heights of 12m and 6m, respectively. Each space is equipped with a full-wall transparent LED screen.(Visual area: $800\,\mathrm{m}^2$)
        & \ChronThumb{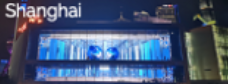} \\
        \addlinespace[4pt]
        \shrinkheight{4pt}
        2020 & \textbf{Guangzhou}\newline
            8K Spring uses the patented technology of curved corners. As the largest 8K naked eye 3D screen in China. 8K Spring combines the innovative application of technology and art such as 8K + naked + AI.
        & \ChronThumb{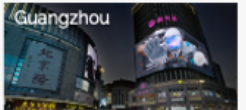} \\
        \addlinespace[4pt]
        \shrinkheight{4pt}
        2020 & \textbf{Chengdu}\newline
            Talkoo Li Mall, Chengdu. Sino-Ocean LED screen display. (Visual area: $426\,\mathrm{m}^2$)
        & \ChronThumb{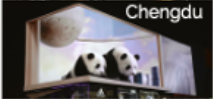} \\
        \addlinespace[4pt]
        \shrinkheight{4pt}
        2020 & \textbf{Foshan}\newline
            Galanz Head Office, Foshan. Galanz's headquarters has created a huge customized exterior curved display to showcase its corporate identity. The urban screen is mounted on the top of its office building and is used to broadcast branded content and innovative custom 3D content. (Visual area:$798\,\mathrm{m}^2$)
        & \ChronThumb{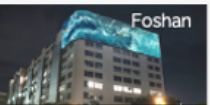} \\
        \addlinespace[4pt]
        \shrinkheight{4pt}

        2021 & \textbf{Xi'an}\newline
            China World Shopping Mall, Xi'an. The screen with 8K resolution, naked-eye 3D video and a 265 degree curved arc display brings a full range of multi-angle visual impact effects to the audience. (Visual area:$1908\,\mathrm{m}^2$)
        & \ChronThumb{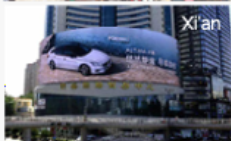} \\
        \addlinespace[4pt]
        \shrinkheight{4pt}
        2021 & \textbf{Beijing}\newline
            Wangfujing Store, Beijing. 4K/8K/8K resolution, naked-eye 3D video. (Visual area:$650\,\mathrm{m}^2$)
        & \ChronThumb{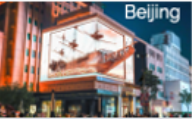} \\
        \addlinespace[4pt]
        \shrinkheight{4pt}
        2021 & \textbf{Guiyang}\newline
            Guiyang city's first naked eye 3D big screen stunning debut Nanning District. (Visual area:$400\,\mathrm{m}^2$)
        & \ChronThumb{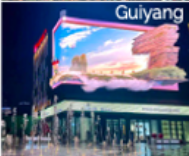} \\
        \addlinespace[4pt]
        \shrinkheight{4pt}
        2021 & \textbf{Fuzhou}\newline
            Dongjiekou, Fuzhou. Fujian's first 8K naked-eye 3D settled in Fuzhou Dongjiekou.
        & \ChronThumb{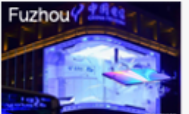} \\
        \addlinespace[4pt]
        \shrinkheight{4pt}
        2021 & \textbf{Shenzhen}\newline
            Shenzhen Radio and Television Tower. Beyond the light of Asia called Asia's largest outdoor 4K large screen. (Visual area:$4501\,\mathrm{m}^2$)
        & \ChronThumb{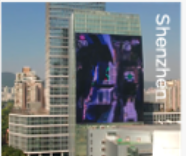} \\
        \addlinespace[4pt]
        \shrinkheight{4pt}

        2022 & \textbf{Ordos}\newline
            Ordos Radio and Television Station is building a naked-eye 3D continuous right angle giant screen with an overall area of $1410\,\mathrm{m}^2$.
        & \ChronThumb{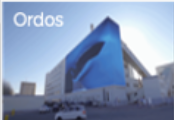} \\
        \addlinespace[4pt]
        \shrinkheight{4pt}
        2022 & \textbf{Nanning}\newline
            Greenland Central Square, Nanning. The largest naked-eye 3D screen in Nanning. (Visual area:$808\,\mathrm{m}^2$)
        & \ChronThumb{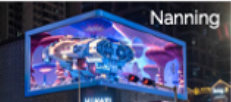} \\
        \addlinespace[4pt]
        \shrinkheight{4pt}
        2022 & \textbf{Qingdao}\newline
            Lanxiu City, Qingdao. Qingdao's first naked-eye 3D special shaped screen. (Visual area:$635\,\mathrm{m}^2$)
        & \ChronThumb{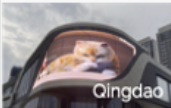} \\
        \addlinespace[4pt]
        \shrinkheight{4pt}
        2022 & \textbf{Zhengzhou}\newline
            China's largest circular screen, Zhengzhou. It is called the first screen in the central plains, and the screen is curved on three sides, with smooth corners. (Visual area:$999\,\mathrm{m}^2$)
        & \ChronThumb{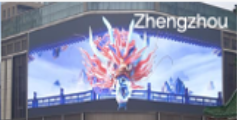} \\
        \addlinespace[4pt]
        \shrinkheight{4pt}
        2022 & \textbf{Chongqing}\newline
            Chongqing Wanda and Meilian. Chongqing, which are nearly 1000\,m\textsuperscript{2} in Wanzhou Wanda Plaza and 2000\,m\textsuperscript{2} in Meilian Plaza.
        & \ChronThumb{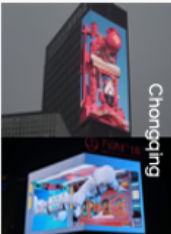} \\
        \addlinespace[4pt]
        \shrinkheight{4pt}
        2023 & \textbf{Lhasa}\newline
            Tianhai international Plaza, Lhasa. 600 m\textsuperscript{2}. The first naked-eye 3D screen in Tibet and the highest naked-eye 3D screen at the current altitude.
        & \ChronThumb{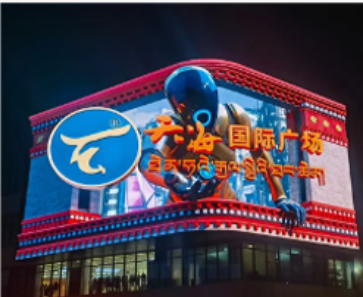} \\
        \addlinespace[4pt]
        \shrinkheight{4pt}
        2024 & \textbf{Tianjin}\newline
            Joy City Shopping Mall. The largest naked-eye 3D screen in North China, presenting a naked-eye 3D audio-visual feast to consumers using ultra-high-definition video technology.(Visual area:$740\,\mathrm{m}^2$)
        & \ChronThumb{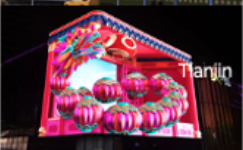} \\
        \addlinespace[4pt]
        \shrinkheight{4pt}
        2024 & \textbf{Suzhou}\newline
            The Light of Suzhou 1088. Suzhou's largest outdoor naked-eye 3D giant screen emerges from the city. (Visual area:$1088\,\mathrm{m}^2$)
        & \ChronThumb{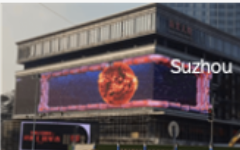} \\
        \addlinespace[4pt]
        \shrinkheight{4pt}
\end{supertabular}
\par
\endgroup


\subsection{Selected Interview Excerpts}
\label{sec:interview-excerpts}

\subsubsection{Interview Overview}
The study included semi-structured interviews with ten participants conducted in 2022--2023. All interviews were conducted in Mandarin Chinese; the selected excerpts below are presented in English translation. Participants' professional backgrounds included architecture, urban planning, government, digital media, manufacturing, and finance. The excerpts are organised by theme and attributed using participant codes. Excerpt identifiers E01--E17 locate individual extracts within this appendix.

\subsubsection{Participant Profiles}
\begin{description}
    \item[A1] International architect.
    \item[A2] Local architect based in Shanghai.
    \item[U1] Urban planning and design professional, Shanghai.
    \item[U2] Urban planning and design professional, Shanghai.
    \item[G1] Senior official, Shanghai Urban Planning Bureau.
    \item[G2] Senior official, third-tier city urban planning bureau.
    \item[D] Digital media strategist, media advertising firm.
    \item[P1] Banking professional, Guanyinqiao District, Chongqing.
    \item[P2] Finance professional, Lujiazui District, Shanghai.
    \item[M] Representative of a Chinese media fa\c{c}ade manufacturer.
\end{description}

\begingroup
\newenvironment{MABInterviewExcerpt}[3]{%
    \begin{list}{}{%
        \setlength{\leftmargin}{0.8em}%
        \setlength{\rightmargin}{0pt}%
        \setlength{\topsep}{4pt}%
        \setlength{\partopsep}{0pt}%
        \setlength{\itemsep}{0pt}%
        \setlength{\parsep}{0pt}%
    }%
    \setlength{\emergencystretch}{1em}%
    \item[]\textbf{[#1] #2}\quad\textnormal{(#3)}\par\nopagebreak[4]
}{%
    \end{list}%
}

\subsubsection{Historical Context and Urban Development}
\begin{MABInterviewExcerpt}{G1}{People's Square as a meeting place}{E01}
‘For us at that time, People's Square was the centre of all centres. In those days without mobile phones, deciding on a clearly defined meeting place was essential to feeling secure.’
\end{MABInterviewExcerpt}

\begin{MABInterviewExcerpt}{U1}{Urban transformation}{E02}
‘During the fifth renovation of People's Square in 2002, the large screen was removed. Along with it went the boutique department store, the music bookstore, the Dalu Hotel, and the residents who lived behind that urban screen. Over the past two decades, the shifts in meeting landmarks have not been solely driven by advances in communication tools—they also reflect the broader transformations taking place in the city of Shanghai.’
\end{MABInterviewExcerpt}

\subsubsection{Commercial and Administrative Aspects}
\begin{MABInterviewExcerpt}{D}{Screen advertising value}{E03}
‘Even though The Citi screen is right next to Aurora, in a similar location and maybe even slightly larger, Aurora's ad fees are still higher because it came first and is now a symbol of Shanghai. It has a special symbolic significance. However, sometimes their ad spaces can be combined for joint screen broadcasts. Beyond the special holiday media shows, some powerful big companies also run joint-screen ads for even greater impact.’
\end{MABInterviewExcerpt}

\begin{MABInterviewExcerpt}{D}{Advertising revenue and circulation}{E04}
‘Aurora's media content is indeed dominated by large commercial advertisements because of their high commercial value. Just imagine, their recent advertising rate is about 600,000 RMB for a continuous hour. With such a prime location, of course they need to profit. Apart from the outdoor ad value itself, you also have to consider the extensive secondary dissemination it creates.’
\end{MABInterviewExcerpt}

\begin{MABInterviewExcerpt}{U1}{Regulatory environment}{E05}
‘Due to government approvals and planning regulations, there won't be any new advertising spaces added in the Bund and Lujiazui areas. In Lujiazui, Aurora Plaza and The Citi's large programmable media façades are the main ones, and no new media façade advertising projects will be approved. This is to protect Lujiazui's nightscape and ensure its rarity.’
\end{MABInterviewExcerpt}

\begin{MABInterviewExcerpt}{G2}{Civic responsibilities}{E06}
‘But besides commercial ads, as a landmark symbol of Shanghai, Aurora must continuously and reliably broadcast the city's official promotional videos and TV station interactive programs— not to mention the city's media shows during holidays and festivals.’
\end{MABInterviewExcerpt}

\subsubsection{Public Reception and Everyday Experience}
\begin{MABInterviewExcerpt}{P2}{Screen interaction}{E07}
‘I don't specifically pay attention to these screens when I'm working. It's usually on weekends, when I happen to pass by with friends, that we notice something special on the screens and then stop to watch for a bit.’
\end{MABInterviewExcerpt}

\begin{MABInterviewExcerpt}{P1}{Everyday encounters}{E08}
‘I work near Guanyin Bridge and see it every day. Sometimes during a break or at lunchtime, I'll sit in that park and watch the screen to pass the time. It's actually more impressive at night than during the day.’
\end{MABInterviewExcerpt}

\begin{MABInterviewExcerpt}{P1}{Sightseeing and screen imagery}{E09}
‘I've also been to places like Liziba Station for sightseeing, and I saw a 3D monorail passing through a building at Guanyin Bridge. It's really interesting, so characteristic of Chongqing. Chongqing truly is an 8D city.’
\end{MABInterviewExcerpt}

\begin{MABInterviewExcerpt}{P1}{Sharing screen experiences}{E10}
‘I remember the Sweater for the Screen—it was so adorable. I took photos and videos and shared them in a group chat with my friends and family. The Super Moon was also really fun! One of my friends even came here that day just to see it. It looked especially beautiful at night.’
\end{MABInterviewExcerpt}

\subsubsection{Urban Planning and Design Perspectives}
\begin{MABInterviewExcerpt}{A2}{International comparisons}{E11}
‘It's very difficult for Chinese architectural firms to produce something like One Shenzhen Bay. If you look at its architecture, landscape, and lighting, they were all done by foreign firms, and they did an excellent job. It's commendable that Shenzhen could realise such a project’
\end{MABInterviewExcerpt}

\begin{MABInterviewExcerpt}{U2}{Regional differences}{E12}
‘Compared to Shanghai, Shenzhen's planning policies are indeed more open and innovative. We often adopt what Shenzhen does well. This difference is due to Shanghai and Shenzhen's distinct strategic positions—Shanghai needs more stable development, while Shenzhen can explore and experiment.’
\end{MABInterviewExcerpt}

\begin{MABInterviewExcerpt}{U1}{Ecological considerations}{E13}
‘It's hard to imagine that in the city centre, where land is so valuable and crowds so dense, there could still be such an ecological wetland. One of Shenzhen's key concepts in achieving 'no disturbance to migratory birds' was to integrate ecological protection into lighting planning.’
\end{MABInterviewExcerpt}

\subsubsection{Industry Insights}
\begin{MABInterviewExcerpt}{M}{Market dynamics}{E14}
‘From an industry perspective, the continuous emergence of these giant urban screens is a kind of industry involution phenomenon. Leveraging their higher-quality media attributes, various media companies are attempting to use these giant screens to grab industry resources and establish absolute market advantages within certain areas.’
\end{MABInterviewExcerpt}

\begin{MABInterviewExcerpt}{A1}{Supply chain evolution}{E15}
‘Our company now uses screen suppliers from mainland China. Initially, we relied on suppliers from Taiwan, but now most are Chinese because the price advantage is just too great.’
\end{MABInterviewExcerpt}

\begin{MABInterviewExcerpt}{U2}{Planning framework}{E16}
‘The design of these city screens generally falls under streetscape planning. There are rarely detailed planning projects, except in certain heritage areas. And even then, it's more about restriction than promotion. Instead of planning where screens should be installed, we focus more on where they must not appear.’
\end{MABInterviewExcerpt}

\begin{MABInterviewExcerpt}{A2}{Technical evolution}{E17}
‘Over the years, I've noticed that screens are getting bigger and their 3D content more abundant. Places like Xujiahui and Wujiaochang have glasses-free 3D screens, curved screens, and screens wrapping around building corners—these are all quite impressive. I myself have never participated in the design of a screen-integrated building or a media building. It seems that, for now, China's design industry still treats the building and its screen as separate elements.’
\end{MABInterviewExcerpt}

\endgroup


\clearpage
\twocolumn[{%
    \begin{minipage}{\textwidth}
        \subsection{Factor-mapping of Chinese urban screens (1990--present)}
        \label{sec:factor-map}
        \centering
        \includegraphics[
            width=\linewidth,
            height=1\textheight,
            keepaspectratio
        ]{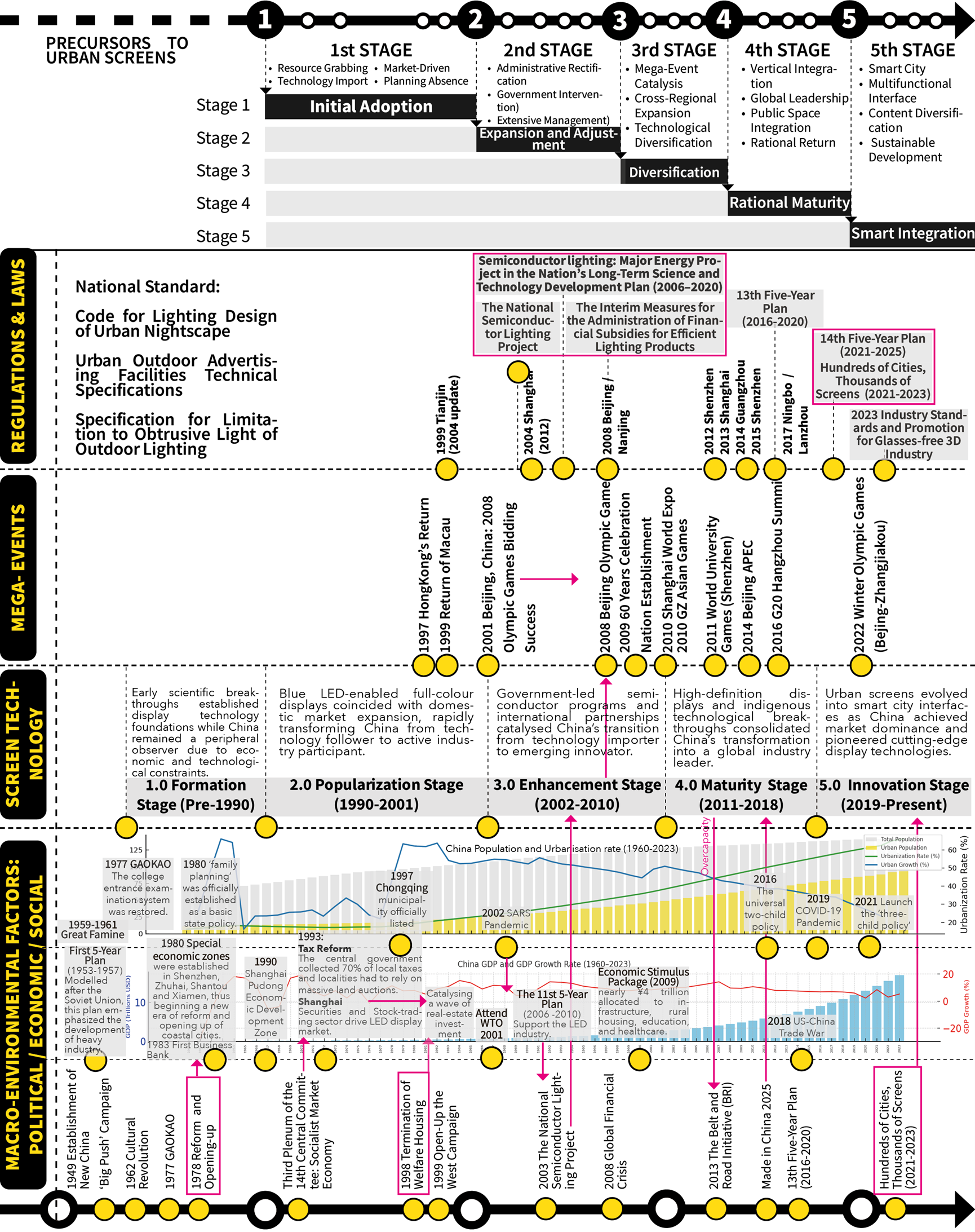}
        \captionsetup{hypcap=false}
        \captionof{figure}{Factor-mapping of Chinese urban screens
        (1990--present). Diagram by the author.}
        \label{map}
        \Description{A timeline linking five stages of urban screen development
        with regulations, major events, screen technology and macro-environmental factors.}
    \end{minipage}%
}]
\clearpage



\subsection{Morphological Inventory of Screen-Based Buildings: Cluster No.~77}
\label{sec:cluster77}

\begin{figure}[H]
    \centering
    \includegraphics[
        width=\linewidth,
        height=0.35\textheight,
        keepaspectratio
    ]{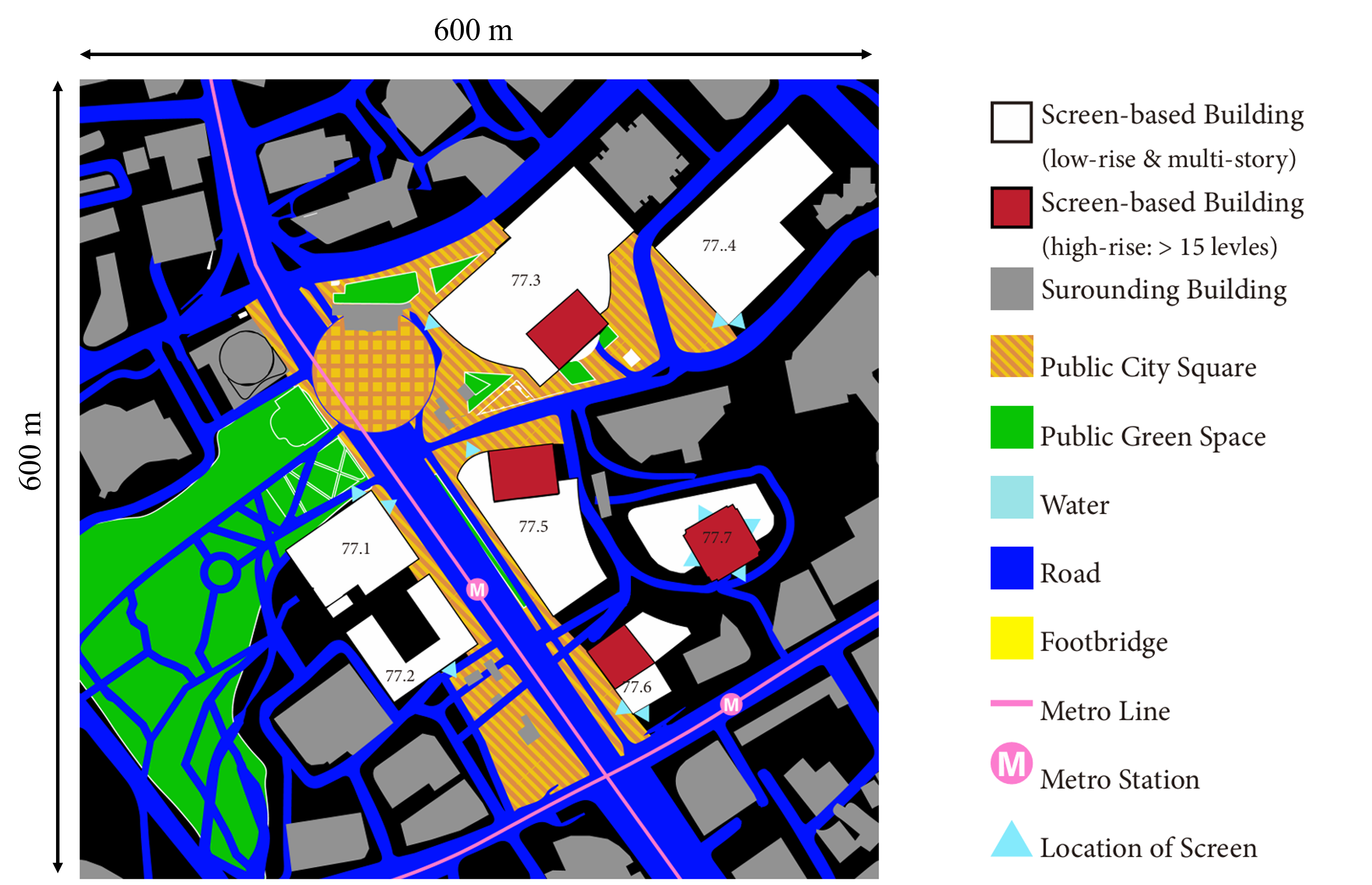}
    \caption{Guanyinqiao screen-based building cluster (No.~77).}
    \label{guanyinqiaoA9}
    \Description{Overview of the Guanyinqiao
    screen-based building cluster.}
\end{figure}

Figure~\ref{guanyinqiaoA9} shows the Guanyinqiao cluster.
Table~\ref{tablec1} brings together each building's image, name, location and recorded screen parameters. All locations are in Jiangbei, Chongqing.

\subsubsection{Abbreviations and Units}
\begingroup
\setlength{\emergencystretch}{1em}
\begin{description}
\item[Identification and installation.]
\textbf{N}: building name;
\textbf{L}: location;
\textbf{P}: building use (property); in the table, Com. means commercial and Mall means shopping mall;
\textbf{IM}: installation method
(\textbf{A}: attached; \textbf{M}: mounted; \textbf{U}: unibody).

\item[Dimensions and operation.]
\textbf{VA}: visual area ($\mathrm{m}^2$); where dimensions are given, they are in $\mathrm{m}\times\mathrm{m}$;
\textbf{VD}: viewing distance (m);
\textbf{DT}: display time;
\textbf{H}: height above ground (m).
\textbf{VF/TF}: visitor flow ($10^4$ passengers/day) / traffic flow ($10^4$ trips/day).

\item[Screen technology.]
\textbf{SL}: light-source type, with resolution where recorded (pixels $\times$ pixels);
\textbf{LDP}: LED display panel;
\textbf{MF}: programmable media fa\c{c}ade building;
\textbf{AP}: architectural projection.

\item[Media content and form.]
\textbf{C/F}: media content / media form;
\textbf{AD}: commercial advertising;
\textbf{PI}: public information;
\textbf{AE}: arts and entertainment;
\textbf{SI}: static images;
\textbf{VO}: video;
\textbf{3D}: glasses-free 3D effect.
\end{description}
\endgroup

\begin{table}[tp]
    \caption{Guanyinqiao screen-based building cluster: project inventory and screen parameters. All locations are in Jiangbei, Chongqing.}
    \label{tablec1}
    \label{tablec1-parameters}
    \centering
    \small
    \setlength{\tabcolsep}{2pt}
    \renewcommand{\arraystretch}{1.08}
    \newcommand{\GQproject}[3]{%
        \textbf{#1}\newline
        #2\newline
        \includegraphics[width=1.25cm,height=1.00cm,keepaspectratio]{#3}%
    }
    \begin{tabular}{@{}
        >{\raggedright\arraybackslash}p{\dimexpr\columnwidth-5.03cm-10\tabcolsep\relax}
        >{\raggedright\arraybackslash}p{0.85cm}
        >{\raggedright\arraybackslash}p{1.02cm}
        >{\raggedright\arraybackslash}p{0.98cm}
        >{\centering\arraybackslash}p{0.88cm}
        >{\raggedright\arraybackslash}p{1.30cm}
        @{}}
        \toprule
        \textbf{Project}\newline\textbf{(N / L / image)}
        & \textbf{P}\newline\textbf{IM}
        & \textbf{VA}\newline\textbf{VD}
        & \textbf{DT}\newline\textbf{H}
        & \textbf{VF/TF}
        & \textbf{C/F}\newline\textbf{SL} \\
        \midrule
        \GQproject{77.1 Carnival Building (3788 Light of Asia)}{SuNing, Guanyinqiao Street}{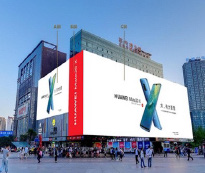}
        & Com.\newline IM: M
        & A: 1045\newline B: 123\newline C: 2586\newline VD:\newline 1000
        & 08:30--\newline 22:00\newline H: 6
        & 120/37
        & AD/\allowbreak PI/\allowbreak AE,\newline 3D/\allowbreak VO\newline SL: LDP\newline (8k) \\
        \midrule
        \GQproject{77.2 Centure Xindu}{9 Guanyinqiao Street}{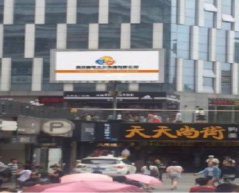}
        & Mall\newline IM: A
        & 52\newline VD:\newline 100
        & 09:00--\newline 22:00\newline H: 8
        & 100/37
        & AD,\newline SI/\allowbreak VO\newline SL: LDP \\
        \midrule
        \GQproject{77.3 Guanyinqiao Imix Park}{8 Guanyinqiao Street}{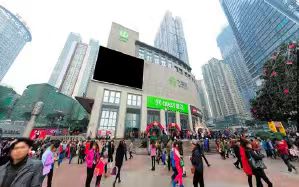}
        & Mall\newline IM: A
        & 213\newline VD:\newline 300
        & 08:00--\newline 21:00\newline H: 11
        & 100/37
        & AD/\allowbreak AE,\newline SI/\allowbreak VO\newline SL: LDP \\
        \midrule
        \GQproject{77.4 Sunshine Mall Outlets}{3 Jianxin N Rd}{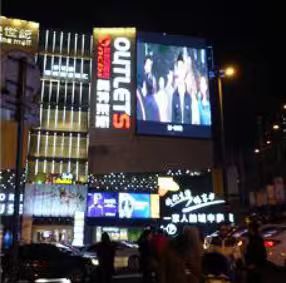}
        & Mall\newline IM: A
        & 100\newline VD:\newline 300
        & 08:00--\newline 22:00\newline H: 6.5
        & 50/50
        & AD,\newline SI/\allowbreak VO\newline SL: LDP \\
        \midrule
        \GQproject{77.5 New World Department Store}{6 Guanyinqiao Street}{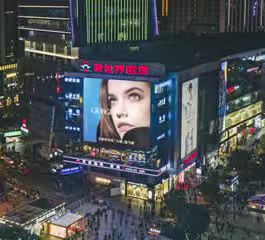}
        & Mall\newline IM: M
        & 440\newline VD:\newline 500
        & 08:30--\newline 22:00\newline H: 11
        & 70/70
        & AD/\allowbreak AE,\newline SI/\allowbreak VO\newline SL: LDP \\
        \midrule
        \GQproject{77.6 Herong Yingjia Centre}{2 Guanyinqiao Street}{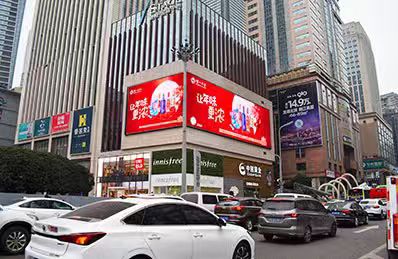}
        & Com./\newline office\newline IM: A
        & 479\newline VD:\newline 500
        & 08:00--\newline 22:00\newline H: 18.4
        & 50/50
        & AD/\allowbreak PI/\allowbreak AE,\newline SI/\allowbreak VO\newline SL: LDP \\
        \midrule
        \GQproject{77.7 CIC International}{10 Guanyinqiao Street}{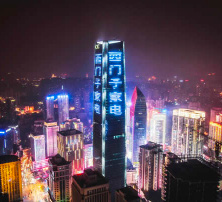}
        & Office\newline IM: U
        & 216004\newline VD:\newline 1000
        & 20:00--\newline 23:00\newline H: 180
        & 50/100
        & AD/\allowbreak PI/\allowbreak AE,\newline SI/\allowbreak VO,\newline Light\newline SL: MF \\
        \bottomrule
    \end{tabular}
    \par\smallskip
    \begin{minipage}{\columnwidth}
        \footnotesize\raggedright
        P: Com. = commercial; Mall = shopping mall; Com./office = commercial and office. Other abbreviations and units are defined in the accompanying text.
    \end{minipage}
\end{table}


\clearpage
\subsection{List of Light of Asia Media Content Projects}
\label{sec:light-of-asia-projects}

\par
\begingroup
\small
\setlength{\tabcolsep}{3pt}
\setlength{\emergencystretch}{1em}
\renewcommand{\arraystretch}{1.10}

\topcaption{Selected media content projects presented on the Light of Asia screen.\label{tableb1}}
\tablefirsthead{%
    \toprule
    \textbf{Project and image} & \textbf{Summary} \\
    \midrule
}
\tablehead{%
    \multicolumn{2}{@{}l@{}}{\small\itshape Table~\ref{tableb1} (continued)} \\[3pt]
    \toprule
    \textbf{Project and image} & \textbf{Summary} \\
    \midrule
}
\tabletail{\bottomrule}
\tablelasttail{\bottomrule}

\begin{supertabular}{@{}
    >{\raggedright\arraybackslash}p{2.25cm}
    >{\raggedright\arraybackslash}p{\dimexpr\columnwidth-2.25cm-2\tabcolsep\relax}
    @{}}
    \textbf{Track Across Building}\newline
    DIGMEGA\newline
    October 2020\newline
    \includegraphics[width=1.65cm,height=1.25cm,keepaspectratio]{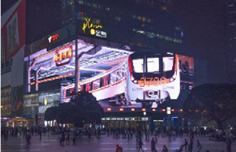}
    & \textbf{Theme.} Urban features, Transportation (Local Cultural, Technological City Branding)\par
      \textbf{Purpose.} Showcase Chongqing's iconic monorail system and unique architecture; Promote city image\par
      \textbf{Presentation and interaction.} Using glasses-free 3D technology; Encouraging the audience to share photos and videos on social media for secondary dissemination and to amplify its impact \\
    \addlinespace[5pt]
    \shrinkheight{5pt}

    \textbf{Spaceship}\newline
    DIGMEGA\newline
    November 2020\newline
    \includegraphics[width=1.65cm,height=1.25cm,keepaspectratio]{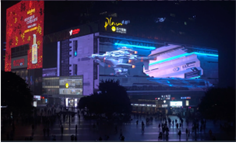}
    & \textbf{Theme.} Technology and Space Exploration (Futuristic, Educational)\par
      \textbf{Purpose.} Awaken interest in science and the future; Enhance landmark status\par
      \textbf{Presentation and interaction.} Glasses-free 3D; encourage social sharing \\
    \addlinespace[5pt]
    \shrinkheight{5pt}

    \textbf{Claw Machine}\newline
    DIGMEGA\newline
    December 2021\newline
    \includegraphics[width=1.65cm,height=1.25cm,keepaspectratio]{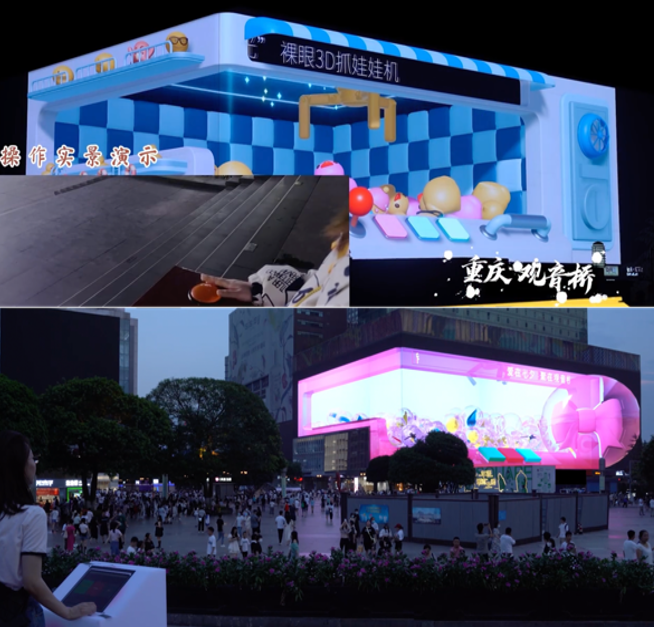}
    & \textbf{Theme.} Interactive Game (User-controlled Claw)\par
      \textbf{Purpose.} Provide innovative and engaging interactive experiences; Chinese Valentine's Day Game\par
      \textbf{Presentation and interaction.} Scene 1: Players control a joystick to grab dolls appearing on the screen, which can be redeemed for corresponding prizes. Scene 2: For Happy Chinese Valentine's Day (August), use this Claw Machine to provide people with a sweet and romantic visual and gaming experience on Valentine's Day. \\
    \addlinespace[5pt]
    \shrinkheight{5pt}

    \textbf{The Light of Asia 3rd Anniversary}\newline
    DIGMEGA\newline
    March 2022\newline
    \includegraphics[width=1.65cm,height=1.25cm,keepaspectratio]{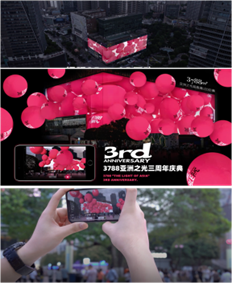}
    & \textbf{Theme.} Cultural; Anniversary Celebration; AR\par
      \textbf{Purpose.} Commemorate the screen's 3rd anniversary; showcase AR capabilities\par
      \textbf{Presentation and interaction.} Together with a glasses-free 3D artwork, a special AR app lets viewers scan around the giant screen to see countless red balls "break out" into the street, extending digital art into real space. \\
    \addlinespace[5pt]
    \shrinkheight{5pt}

    \textbf{Super Moon + AR Constellations}\newline
    DIGMEGA\newline
    June/July 2022\newline
    \includegraphics[width=1.65cm,height=1.25cm,keepaspectratio]{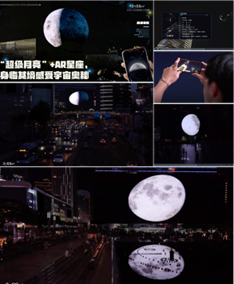}
    & \textbf{Theme.} An immersive experience of the mysteries of the universe. (Astronomy, Nature)\par
      \textbf{Purpose.} Provide an immersive astronomical experience showcasing the Super Moon and constellations. Educate about celestial events, Engage public interest.\par
      \textbf{Presentation and interaction.} In 2022, the largest full moon of the year, the Super Moon, appeared in the night sky. The surface of the moon was vividly displayed, accompanied by AR effects presenting constellations with phones, creating a rare astronomical spectacle at Guanyinqiao. \\
    \addlinespace[5pt]
    \shrinkheight{5pt}

    \textbf{Elephant Music Fountain}\newline
    DIGMEGA\newline
    April 2024\newline
    \includegraphics[width=1.65cm,height=1.25cm,keepaspectratio]{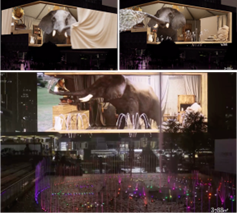}
    & \textbf{Theme.} Elephants playing with a music fountain.\par
      \textbf{Purpose.} Create a captivating visual; Promote local attractions.\par
      \textbf{Presentation and interaction.} The elephant interacts with the Guanyinqiao music fountain in a playful water dance, presenting a new approach to outdoor digital media and scene-based marketing. Combining visual and environmental elements, encouraging viewers to immerse themselves in this visual adventure that blends physical surroundings with digital artistry. \\
    \addlinespace[5pt]
    \shrinkheight{5pt}

    \textbf{Sweater for the Screen}\newline
    DIGMEGA\newline
    December 2023\newline
    \includegraphics[width=1.65cm,height=1.25cm,keepaspectratio]{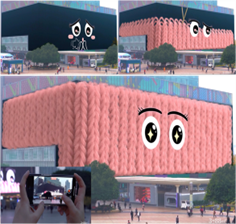}
    & \textbf{Theme.} Screen wearing a sweater.\par
      \textbf{Purpose.} Brings a playful and festive atmosphere to the city. It also has a welfare component, reminding audiences to stay warm during the cold weather.\par
      \textbf{Presentation and interaction.} On the day of the Major Snow solar term, the screen was dressed in a sweater using creative 3D visuals, adding warmth to the city during the winter season. This engaging content immediately attracted many residents and tourists to visit and capture the moment, encouraging physical interaction through on-site visits and social sharing. \\
    \addlinespace[5pt]
    \shrinkheight{5pt}

    \textbf{Lightscales}\newline
    DIGMEGA $\times$ Mario Carrillo\newline
    July 2024\newline
    \includegraphics[width=1.65cm,height=1.25cm,keepaspectratio]{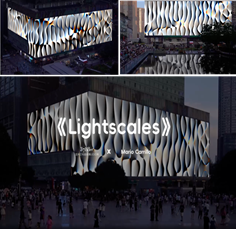}
    & \textbf{Theme.} (Light, Abstract art)\par
      \textbf{Purpose.} Explore light manipulation in digital art, Create visual wonder.\par
      \textbf{Presentation and interaction.} Screens can be seen as portals: windows into alternate worlds. By experimenting with light and abstract forms, this project creates a captivating visual experience. The interaction method invites audiences to immerse themselves in the dynamic interplay of light, engaging with the piece on a sensory level and experiencing the wonder and aesthetics of light in an entirely new digital format. \\
\end{supertabular}
\par
\endgroup

\end{document}